\pdfoutput=1
\documentclass{article} 
\usepackage{iclr2021_conference,times}
\iclrfinalcopy

\usepackage[utf8]{inputenc} 
\usepackage[T1]{fontenc}    
\usepackage{hyperref}       
\usepackage{url}            
\usepackage{booktabs}       
\usepackage{amsfonts}       
\usepackage{nicefrac}       
\usepackage{microtype}      

\usepackage{times}
\usepackage{epsfig}
\usepackage{graphicx}
\usepackage{amsmath}
\usepackage{amssymb}
\usepackage{paralist}
\usepackage{dsfont}
\usepackage{multirow}
\usepackage{wrapfig}
\usepackage{xcolor}
\usepackage{subfigure}
\usepackage{pifont}
\usepackage{lipsum}
\input{taesup.sty}

\newcommand{\cmark}{\ding{51}}%
\newcommand{\xmark}{\ding{55}}%

\title{GAN2GAN: Generative Noise Learning for Blind Denoising with Single Noisy Images }

\author{%
  Sungmin Cha$^1$, Taeeon Park$^1$, Byeongjoon Kim$^2$, Jongduk Baek$^2$ and Taesup Moon$^{3}$\thanks{Corresponding author (E-mail: \texttt{tsmoon@snu.ac.kr})}\\
  Sungkyunkwan University$^1$, Yonsei University$^2$, Seoul National University$^3$, South Korea \\
  \texttt{\{csm9493,pte1236\}@skku.edu}, \texttt{bjkim2006@naver.com},\\ \texttt{jongdukbaek@yonsei.ac.kr},
  \texttt{tsmoon@snu.ac.kr}
}

\begin{document}

\maketitle

\begin{abstract}

We tackle a challenging blind image denoising problem, in which only single distinct noisy images are available for training a denoiser, and no information about noise is known, except for it being zero-mean, additive, and independent of the clean image. In such a setting, which often occurs in practice, it is not possible to train a denoiser with the standard discriminative training or with the recently developed Noise2Noise (N2N) training; the former requires the underlying clean image for the given noisy image, and the latter requires two independently realized noisy image pair for a clean image. To that end, we propose GAN2GAN (Generated-Artificial-Noise to Generated-Artificial-Noise) method that first learns a generative model that can 1) simulate the noise in the given noisy images and 2) generate a rough, noisy estimates of the clean images, then 3) iteratively trains a denoiser with subsequently synthesized noisy image pairs (as in N2N), obtained from the generative model. In results, we show the denoiser trained with our GAN2GAN achieves an impressive denoising performance on both synthetic and real-world datasets for the blind denoising setting; it almost approaches the performance of the standard discriminatively-trained or N2N-trained models that have more information than ours, and it significantly outperforms the recent baseline for the same setting, \textit{e.g.}, Noise2Void, and a more conventional yet strong one, BM3D. \textcolor{black}{The official code of our method
 is available at \href{https://github.com/csm9493/GAN2GAN}{https://github.com/csm9493/GAN2GAN}.}


\end{abstract}

\pdfoutput=1

\section{Introduction}



Image denoising is one of the oldest problems in image processing and low-level computer vision, yet it still attracts lots of attention due to the fundamental nature of the problem. A vast number of algorithms have been proposed over the past several decades, and recently, the CNN-based methods, \textit{e.g.}, \cite{cha2018fully, ZhaZuoCheMenZha17, TaiYanLiuXu17, LieWenFanLoyHua18}, 
became the throne-holders in terms of the PSNR performance. 
The main approach of the most CNN-based denoisers is to apply the discriminative learning framework with (clean, noisy) image pairs and \emph{known} noise distribution assumption. 
While being effective, such framework also possesses a couple of limitations that become critical in practice; the assumed noise distribution may be mismatched to the actual noise in the data or obtaining the noise-free clean target images is not always possible or very expensive, \emph{e.g.}, medical imaging (CT or MRI) or astrophotographs.



Several attempts have been made to resolve above issues. For the noise uncertainty, the so-called \emph{blind training} have been proposed. Namely,
a denoiser can be trained with a composite training set that contains images corrupted with multiple, pre-defined noise levels or distributions,
and such blindly trained denoisers, \textit{e.g.}, DnCNN-B in \cite{ZhaZuoCheMenZha17}, were shown to alleviate the mismatch scenarios to some extent.
However, the second limitation, \emph{i.e.}, the requirement of clean images for building the training set, still remains. As an attempt to address this second limitation, \cite{lehtinen2018noise2noise} recently proposed the Noise2Noise (N2N) method. It has been shown that a denoiser, which has a negligible performance loss, can be trained without the clean target images, as long as two independent noisy image realizations for the same underlying clean image are available. Despite its effectiveness, the requirement of the \emph{two} independently realized noisy image pair for a single clean image, which may hardly be available in practice, is a critical limiting factor for N2N. 


In this paper, we consider a setting in which neither of above approach is applicable, 
namely, the pure \textit{unsupervised} blind denoising setting where only \emph{single} distinct noisy images are available for training.
Namely, nothing is known about the noise other than it being zero-mean, additive, and independent of the clean image, and neither the clean target images for blind training nor the noisy image pairs for N2N training is available. While some recent work, \textit{e.g.}, \cite{krull2018noise2void,batson2019noise2self,(HQDenoising)laine2019high}, took the self-supervised learning (SSL) approach for the same setting, we take a generative learning approach. The crux of our method is to first learn a Wasserstein GAN \citep{arjovsky2017wasserstein}-based generative model that can 1) learn and simulate the noise in the given noisy images and 2) generate rough, initially denoised images. 
Using such generative model, we then synthesize noisy image pairs by corrupting each of the initially denoised images with the simulated noise \textit{twice} and use them to train a CNN denoiser as in the N2N training (\textit{i.e.}, Noisy N2N).  
We further show that \textit{iterative} N2N training with refined denoised images can 
significantly improve the final denoising performance. 
We dubbed our method as GAN2GAN (Generated-Artifical-Noise to Generated-Artificial-Noise) and show that the denoiser trained with our method can 
achieve (sometimes, even outperform) the performance of the standard supervised-trained or N2N-trained blind denoisers for the white Gaussian noise case. Furthermore, for mixture/correlated noise or real-world noise in microscopy/CT images, for which the exact distributions are hard to know \textit{a priori}, we show our denoiser significantly outperforms those standard blind denoisers, which are mismatch-trained with white Gaussian noise, as well as other baselines that operate in the same condition as ours: the SSL baseline, N2V \citep{krull2018noise2void}, and a more conventional BM3D \citep{bm3d}.

\section{Related Work}

Several works have been proposed to overcome the limitation of the vanilla supervised learning based denoising. As mentioned above, Noise2Self (N2S) \citep{batson2019noise2self} and Noise2Void (N2V) \citep{krull2018noise2void} recently applied self-supervised learning (SSL) approach to train a denoiser only with single noisy images. Their settings exactly coincide with ours, but we show later that our GAN2GAN significantly outperforms them. More recently, \cite{(HQDenoising)laine2019high} improved N2V by incorporating specific noise likelihood models with Bayesian framework, however, their method \textit{required} to know the exact noise model and could not be applied to  more general, unknown noise settings. Similarly, \cite{(Sure)soltanayev2018training} proposed SURE (Stein's Unbiased Risk Estimator)-based denoiser that can also be trained with single noisy images, but it worked only with the \textit{Gaussian} noise. Their work was extended in \cite{(eSure)zhussip2019extending}, but it required noisy image \textit{pairs} as in N2N as well as the Gaussian noise constraint. 
\cite{chen2018image} devised GCBD method to learn and generate noise in the given noisy images using W-GAN \cite{arjovsky2017wasserstein} and utilized the unpaired clean images to build a supervised training set. Our GAN2GAN is related to \cite{chen2018image}, but we significantly improve their noise learning step and do \emph{not} use the clean data at all. 
Table \ref{table:comparison} summarizes and compares the settings among the above mentioned recent baselines. We clearly see that only our GAN2GAN and N2V do not utilize any ``sidekicks'' that other methods use. 

\vspace{-.2in}
\begin{table}[h]\caption{Summary of different settings among the recent baselines.}
\centering
\smallskip\noindent
\resizebox{0.6\linewidth}{!}{
\begin{tabular}{|c||c|c|c|}
\hline
                             
Alg.$\backslash$ Requirements & Clean image  & Noisy ``pairs'' & Noise model\\\hline \hline     
N2N [\cite{lehtinen2018noise2noise}] & \xmark& \cmark & \xmark\\ \hline
HQ SSL [\cite{(HQDenoising)laine2019high}] &\xmark &\xmark & \cmark  \\\hline
SURE [\cite{(Sure)soltanayev2018training}]  & \xmark&\xmark & \cmark\\ \hline
Ext. SURE [\cite{(eSure)zhussip2019extending}]                 &       \xmark       &    \cmark  &  \cmark         \\         \hline
GCBD [\cite{chen2018image}]                   &      \cmark      &   \xmark   & \xmark  \\            \hline
\hline
N2V [\cite{krull2018noise2void}] & \xmark &\xmark &\xmark  \\ \hline
GAN2GAN (Ours) & \xmark &\xmark &\xmark \\ \hline
\end{tabular}
}
    \label{table:comparison}
\end{table}

\textcolor{black}{Additionally, there are recently published papers on blind image denoising but these also have a difference with ours.  
\cite{anwar2019real, zhang2018residual} suggest effective CNN architectures for denoising, however, they only consider the setting in which clean images are necessary for training.
\cite{zamir2020cycleisp} considers the denoising of specific camera settings, and it also requires clean sRGB images as well as the knowledge of the noise level. Thus, it cannot be applied to the complete blind setting as ours, in which no information on the specific noise distribution or clean images is available.}


More classical denoising methods are capable of denoising solely based on the single noisy images by applying various principles, \textit{e.g.}, filtering-based \cite{Buadasetal05,bm3d}, optimization-based \cite{ElaAha06,mai09}, Wavelet-based \cite{DonJoh95}, and effective prior-based \cite{ZorWei11}. Those methods typically are, however, computationally intensive during the inference time and cannot be \textit{trained} from a separate set of noisy images, which limits their denoising performance. 
Another line of recent work worth mentioning is the deep learning-based priors or regularizers,
\textit{e.g.}, \cite{UlyVedLem18,YehLimChen18,LunzOktemSch18}, but their PSNRs still fell short of the supervised trained CNN-based denoisers. 
\section{Motivation}\label{sec:motivation}





In order to develop the core intuition for motivating our method, we first consider a simple, single-letter Gaussian noise setting.
Let 
$Z=X+N$ 
be the noisy observation of $X\sim\Ncal(0,\sigma_X^2)$, corrupted by the $N\sim\Ncal(0,\sigma_N^2)$.
It is well known that the minimum MSE (MMSE) estimator of $X$ given $Z$ is 
$f^*_{\text{MMSE}}(Z)=\mathbb{E}(X|Z)=\frac{\sigma_X^2}{\sigma_X^2+\sigma_N^2}Z. 
$ 
We now identify the optimality of N2N in this setting. 

\noindent\textbf{N2N} \ \ Assume that we have two i.i.d. copies of the noise $N$: $N_1$ and $N_2$. Then, let 
$Z_1 = X+N_1$ and $Z_2=X+N_2$ be the two independent noisy observation pairs of $X$. The N2N in this setting corresponds to obtaining the MMSE estimator of $Z_2$ given $Z_1$, 
\begin{eqnarray}
f_{\text{N2N}}(Z_1)\triangleq\arg\min_{f}\mathbb{E}(Z_2-f(Z_1))^2 = \mathbb{E}(Z_2|Z_1) = \mathbb{E}(X+N_2|Z_1) \stackrel{(a)}{=} \mathbb{E}(X|Z_1) = \frac{\sigma_X^2}{\sigma_X^2+\sigma_N^2}Z_1,\label{eq:n2n}
\end{eqnarray}
in which (a) follows from $N_2$ being independent of $Z_1$. Note (\ref{eq:n2n}) has the exact same form as $f^*_{\text{MMSE}}(Z)$, hence, estimating $X$ with $f_{\text{N2N}}(Z)$ also achieves the MMSE, in line with \citep{lehtinen2018noise2noise}.


\noindent\textbf{``Noisy'' N2N} \ \  Now, consider the case in which we again have the two i.i.d. $N_1$ and $N_2$, but the noisy observations are of a \textit{noisy} version of $X$. Namely, let $X'=X+N_0$, in which $N_0\sim\Ncal(0,\sigma_0^2)$, and denote $Z'_1=X'+N_1$ and $Z'_2=X'+N_2$ as the noisy observation pairs. Then, we can define a ``Noisy'' N2N estimator as the MMSE estimator of $Z'_2$ given $Z'_1$,
\begin{eqnarray}
f_{\text{Noisy N2N}}(Z'_1,y) \triangleq \arg\min_f \mathbb{E}(Z_2'-f(Z_1'))^2 = \mathbb{E}(X'|Z_1') = \frac{\sigma_X^2(1+y)}{\sigma_X^2(1+y)+\sigma_N^2}Z'_1,\label{eq:noisy_n2n}
\end{eqnarray}
in which we denote $y\triangleq \sigma_0^2/\sigma_X^2$ and assume that $0\leq y <1$.  Note clearly (\ref{eq:noisy_n2n}) coincides with (\ref{eq:n2n}) when $y=\sigma_0^2=0$. Following N2N, (\ref{eq:noisy_n2n}) is essentially estimating $X'$ based on $Z'=X'+N$. An interesting subtle question is what happens when we use the mapping $f_{\text{Noisy N2N}}(Z,y)$ for estimating $X$ given $Z=X+N$, \textit{not} $X'$ given $Z'$. Our theorem below, of which proof is in the Supplementary Material (S.M.), shows that for a sufficiently large $\sigma_0^2$, $f_{\text{Noisy N2N}}(Z,y)$ gives a better estimate of $X$ than $X'$.
\vspace{-.05in}
\begin{theorem}\label{thm:mse}
Consider the single-letter Gaussian setting and 
$f_{\text{Noisy N2N}}(Z,y)$ obtained in (\ref{eq:noisy_n2n}). Also, assume  $0<y<1$. Then, there exists some $y_0$ s.t. $\forall y\in(y_0,1)$, $\mathbb{E}(X-f_{\text{Noisy N2N}}(Z,y))^2<\sigma_0^2$. 
\end{theorem}\vspace{-.1in}



\begin{wrapfigure}{r}{0.35\textwidth}
\vspace{-.05in}
    \centering
    \includegraphics[width=0.35\textwidth]{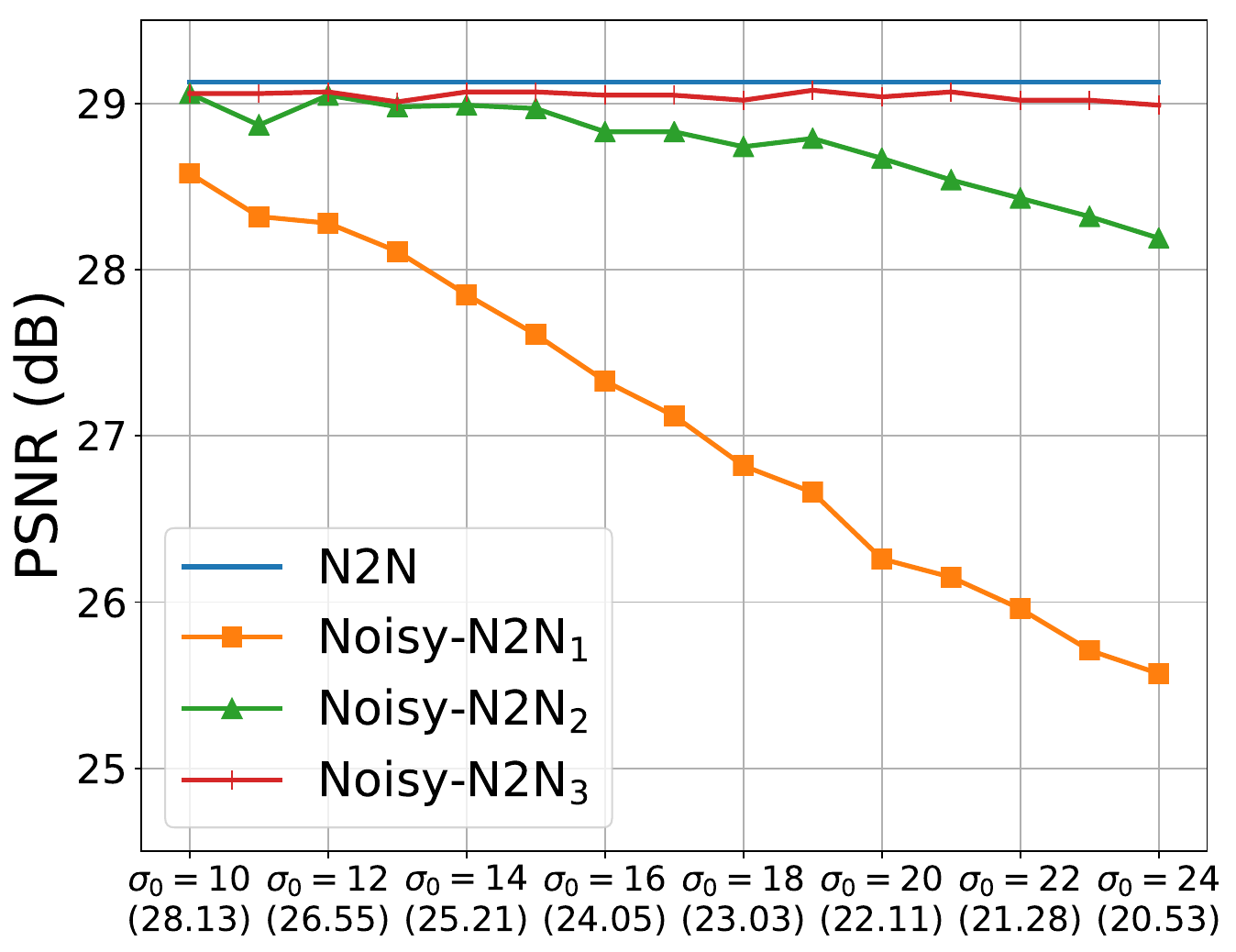}
    \vspace{-.13in}
    \caption{Iterative Noisy N2N.
    }\label{fig:noisy_n2n}
    \vspace{-.13in}
\end{wrapfigure}
Theorem \ref{thm:mse} provides a simple, but useful, intuition that motivates our method;
if simulating the noise in the images is possible, 
we may carry out the N2N training iteratively, provided that a rough \textit{noisy} estimate of the clean image is initially available. Namely, we can first simulate the noise to generate noisy observation pairs 
of the initial noisy estimate, then do the Noisy N2N training with them to obtain a denoiser that may result in a better estimate of the clean image when applied to the actual noisy image subject to denoising (as in Theorem \ref{thm:mse}).
Then, we can refine the estimates by \textit{iterating} the Noisy N2N training with the generated noisy observation pairs of the previous step's estimate of the clean image, until convergence. 

To check whether above intuition is valid, we carry out a feasibility experiment.
Figure \ref{fig:noisy_n2n} shows the denoising results on BSD68 \citep{foe} for Gaussian noise with $\sigma=25$. The blue line is the PSNR of the N2N model trained with noisy observation pairs of the \textit{clean} images in the BSD training set, serving as an upper bound. The orange line, in contrast, is the PSNR of the Noisy N2N$_1$ model that is trained with the noisy observation pairs of the \textit{noisy} estimates for the clean images, which were set to be another Gaussian noise-corrupted training images. The standard deviations ($\sigma_0$) of the Gaussian for generating the noisy estimates are given in the horizontal axis, and the corresponding PSNRs of the estimates are given in the parentheses. 
Although Noisy N2N$_1$ clearly lies much lower than the N2N upper bound, we note its PSNR is still higher than that of the initial noisy estimates, which is in line with Theorem \ref{thm:mse}. Now, if we iterate the Noisy N2N 
with the previous step's denoised images (\textit{i.e.}, Noisy-N2N$_{2}$/Noisy-N2N$_{3}$ for second/third iterations, respectively), we observe that the PSNR significantly improves and approaches the ordinary N2N for most of the initial $\sigma_0$ values.  
Thus, we observe the intuition from Theorem \ref{thm:mse} generalizes well to the image denoising case in an ideal setting, where the noise can be perfectly simulated, and the initial noisy estimates are  Gaussian corrupted versions. The remaining question is whether we can also obtain similar results for the blind image denoising setting. We show our generative model-based approach in details in the next section. 

\vspace{-.02in}
\section{Main Method: Three Components of GAN2GAN}

\label{sec:gan2gan}


To concretely describe our method, we first set the notations. We assume the noisy image $\mathbf{Z}$ is generated by $\mathbf{Z}=\mathbf{x}+\mathbf{N}$, in which $\mathbf{x}$ denotes the underlying clean image and $\mathbf{N}$ denotes the zero-mean, additive noise that is independent of $\xb$. For training a denoiser, we do not assume either the distribution or the covariance of $\mathbf{N}$ is known. Moreover, we assume only a database of $n$ \textit{distinct} noisy images, $\mathcal{D}=\{\mathbf{Z}^{(i)}\}_{i=1}^n$, is available for learning a denoiser.
A CNN-based denoiser is denoted as $\hat{\mathbf{X}}_{\bm\phi}(\mathbf{Z})$ with $\bm\phi$ being the model parameter, and we use the standard quality metrics, PSNR/SSIM, for evaluation. 
Our method consists of three parts; 1) smooth noisy patch extraction, 2) training a generative model, and 3) iterative GAN2GAN training of $\hat{\mathbf{X}}_{\bm\phi}(\mathbf{Z})$, each of which we elaborate below. 



\vspace{-.02in}
\subsection{Smooth noisy patch extraction}\label{subsec:noisy patch}
\vspace{-.02in}

The first step is to extract the noisy image patches from $\mathcal{D}$ that correspond to smooth, homogeneous areas. 
Our extraction method is similar to that of the GCBD proposed in \citep{chen2018image}, but we make a critical improvement. The GCBD determines a patch $\pb$ (of pre-determined size) is smooth if it satisfies the following for \textit{all} of its smaller sub-patches, $\mathbf{q}_j$, with some hyperparameters $\mu, \gamma\in(0,1)$:
\begin{eqnarray}
|\mathbb{E}(\mathbf{q}_{j})-\mathbb{E}(\pb)|\leq\mu \mathbb{E}(\pb), \ \ 
 |\mathbb{V}(\mathbf{q}_{j})-\mathbb{V}(\pb)|\leq \gamma \mathbb{V}(\pb),
\label{eq:var_smooth}
\end{eqnarray}
in which $\mathbb{E}(\cdot)$ and $\mathbb{V}(\cdot)$ are the empirical mean and variance of the pixel values.

\begin{figure}[h]
\centering  
\subfigure[Histograms of empirical $\hat{\sigma}$.]
{\includegraphics[width=0.3\linewidth]{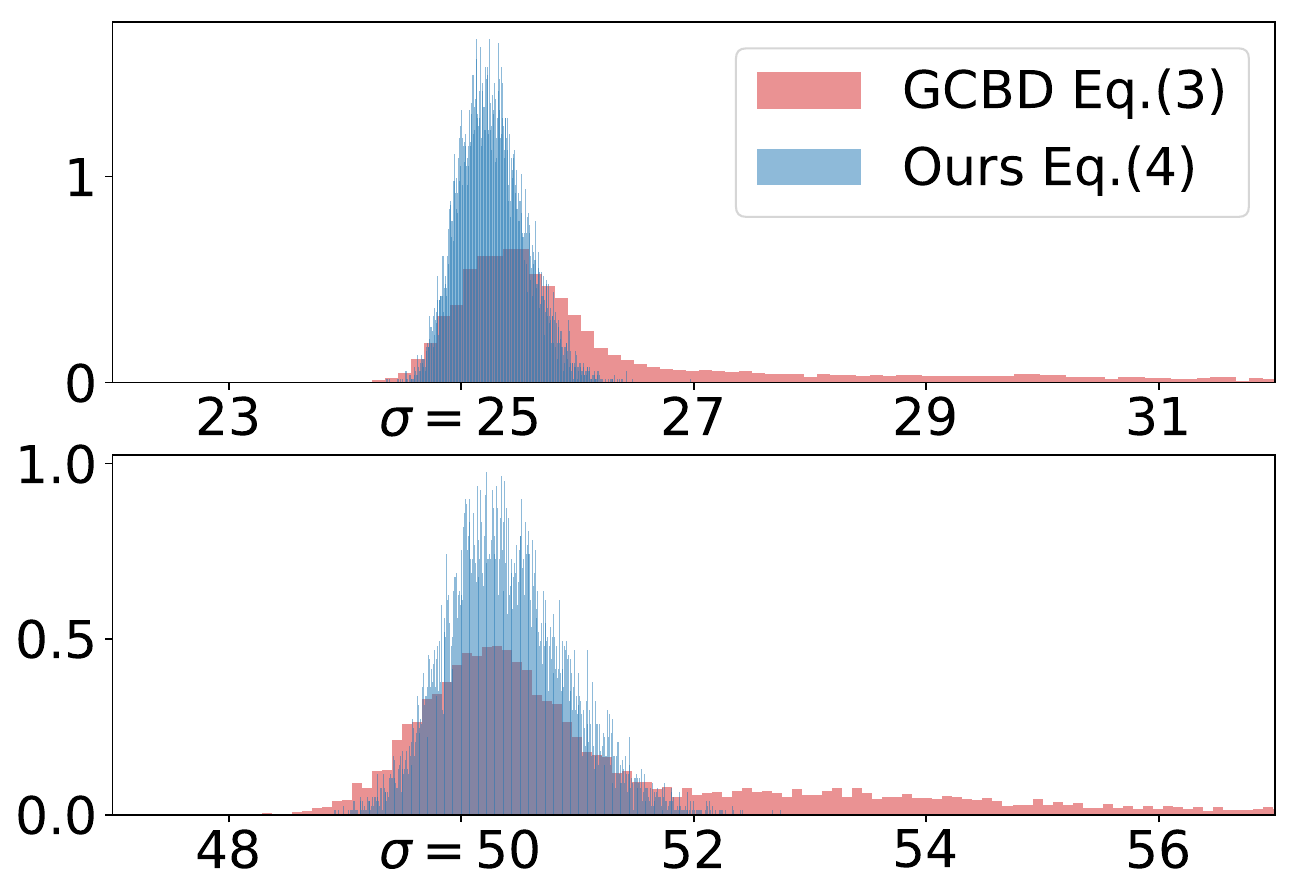}
\label{figure:histogram}}
\hspace{.1in}
\subfigure[Examples of extracted noise patches ($\sigma = 25$)]
{\includegraphics[width=0.45\linewidth]{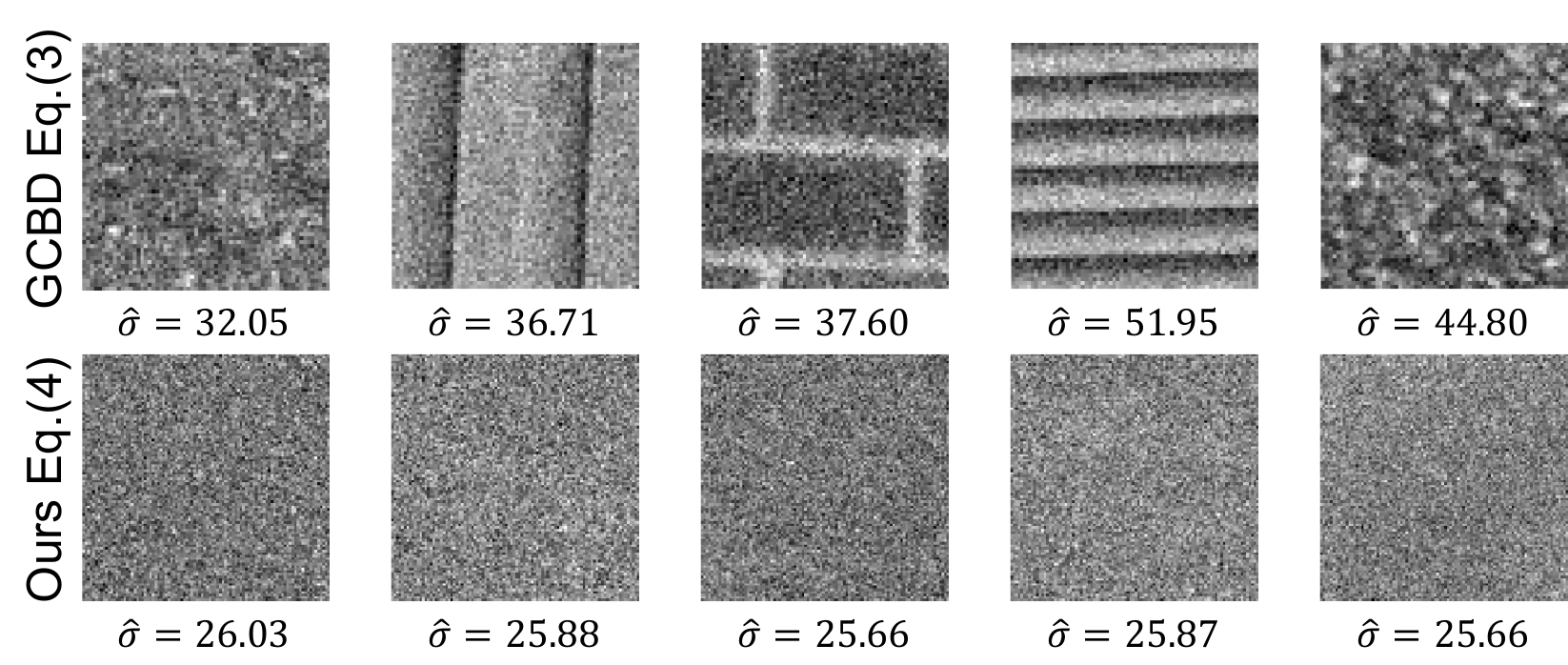}
\label{figure:patches}}
\caption{Comparison of smooth noisy patch extraction rules.}\vspace{-.05in}\label{figure:gcbd_compare}
 \end{figure}

While (\ref{eq:var_smooth}) works for extracting smooth patches to some extent, as we show in Figure \ref{figure:patches}, it 
does not rule out choosing patches with high-frequency repeating patterns, which are far from being smooth. 
Thus, we instead use the 2D discrete wavelet transform (DWT) for a new extraction rule; namely, we determine $\pb$ is smooth if its four sub-band decompositions obtained by DWT, 
$\{W_k(\pb)\}_{k=1}^4$, satisfy
\begin{eqnarray}
\frac{1}{4}\sum_{k=1}^4\Big|\hat{\bm\sigma}(W_k(\pb))-\mathbb{E}[\hat{\bm\sigma}_W(\pb)]\Big|
\leq\lambda\mathbb{E}[\hat{\bm\sigma}_W(\pb)],
\label{eq:dwt_smooth}\vspace{-.05in}
\end{eqnarray}
in which $\hat{\bm\sigma}(\cdot)$ is the empirical standard deviation of the wavelet coefficients, $\mathbb{E}[\hat{\bm\sigma}_W(\pb)]\triangleq\frac{1}{4}\sum_{k=1}^4\hat{\bm\sigma}(W_k(\pb))$, and $\lambda\in(0,1)$ is a hyperparameter.
This rule is much simpler than (\ref{eq:var_smooth}), which has to be evaluated for all the sub-patches, $\{\mathbf{q}_j\}$. 
Once $N$ patches are extracted from $\mathcal{D}$ using (\ref{eq:dwt_smooth}),
we subtract each patch with its mean pixel value, and obtain a set of ``noise'' patches, $\mathcal{N}=\{\nb^{(j)}\}_{j=1}^N$. Such subtraction is valid since all the pixel values should be close to their mean in a smooth patch, and the noise is assumed to be zero-mean, additive.  


Figure \ref{figure:gcbd_compare} compares the rules (\ref{eq:var_smooth}) and (\ref{eq:dwt_smooth}) by showing the quality of the ``noise'' patches extracted from 1,000 Gaussian-corrupted images 
. 
The two plots in Figure \ref{figure:histogram} show the normalized histograms of the empirical standard deviations, $\hat{\sigma}$, of the extracted patches when true $\sigma=\{25,50\}$, respectively. We clearly observe that while the $\hat{\sigma}$'s for (\ref{eq:dwt_smooth}) are mostly concentrated on true $\sigma$, those of  (\ref{eq:var_smooth}) have much higher variation. In addition, Figure \ref{figure:patches} visualizes the randomly sampled patches of which $\hat{\sigma}$'s were above the 90-th percentile among the extracted patches for each rule (when $\sigma=25$). Again, it is obvious that (\ref{eq:var_smooth}) also may result in selecting the patches with high-frequency patterns, whereas (\ref{eq:dwt_smooth}) is much more effective for extracting accurate noise patches. 
Later, we show (in Figure \ref{fig:gcbd_compare_models}) that such improved quality of the noise patches by our  (\ref{eq:dwt_smooth}) plays an \textit{essential} role; namely, our pure unsupervised learning based denoiser using (\ref{eq:dwt_smooth}) even outperforms the clean target image based denoiser in \citep{chen2018image} using (\ref{eq:var_smooth}).

\vspace{-.02in}

\begin{figure*}[t]
\centering
\subfigure[Getting $\mathcal{D}$ \& $\mathcal{N}$]{\includegraphics[width=0.17\linewidth]{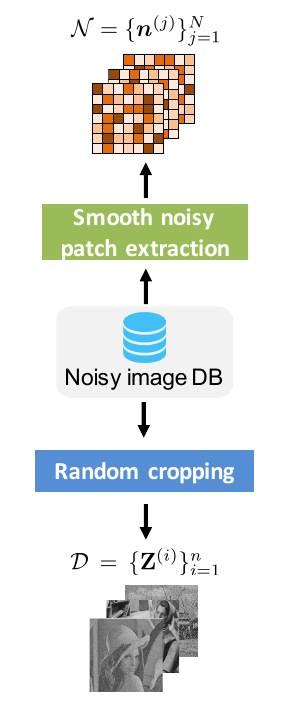}}\hspace{.4in}
\subfigure[The model architecture with three generators and two critics.
]{\includegraphics[width=0.65\linewidth]{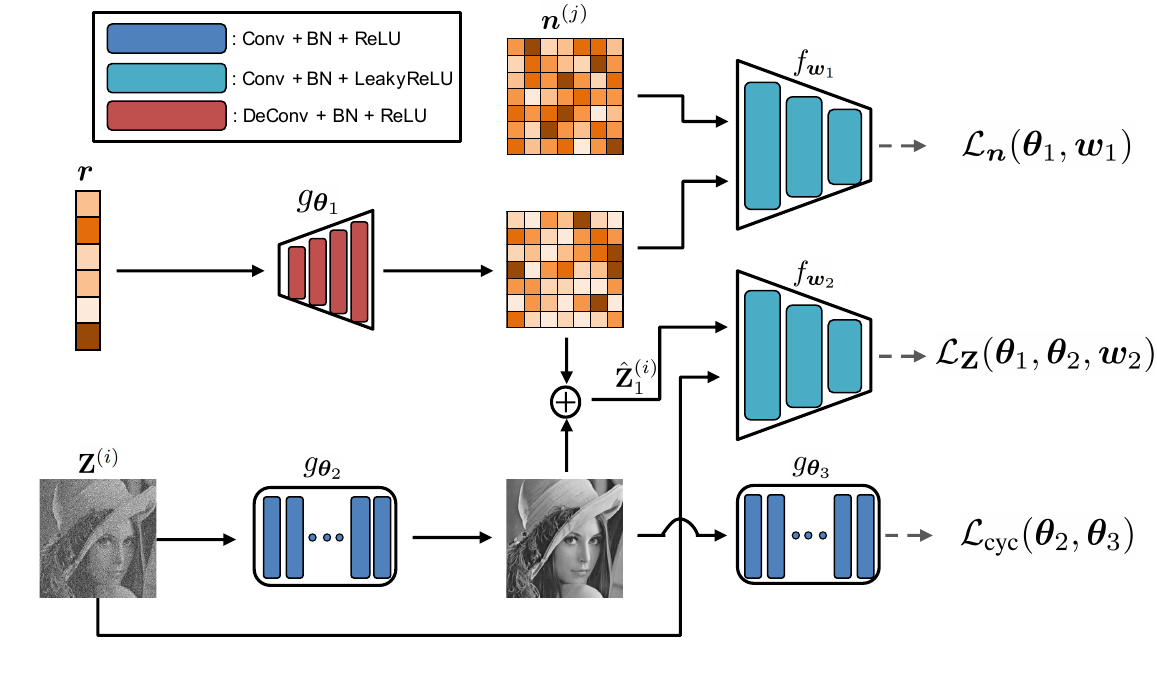}}
\caption{Overall structure of the W-GAN based generative model.}\label{fig:overall model}\vspace{-.15in}
\end{figure*}
\subsection{Training a W-GAN based generative model}\label{subsec:gan_training}
\vspace{-.02in}

Equipped with $\mathcal{D}=\{\Zb^{(i)}\}_{i=1}^n$ and the extracted noise patches $\mathcal{N}=\{\nb^{(j)}\}_{j=1}^N$, we train a generative model, 
which can learn and simulate the noise as well as generate initial noisy estimates of the clean images, hence, realize the Noisy N2N training explained in 
Section \ref{sec:motivation}.
As shown in Figure \ref{fig:overall model}, our model has three generators, $\{g_{\thetab_1},g_{\thetab_2},g_{\thetab_3}\}$, and two critics, $\{f_{\wbb_1},f_{\wbb_2}\}$, in which the subscripts stand for the model parameters. 
The loss functions associated with the components of our model are:
\begin{align}
\mathcal{L}_{\nb}(\thetab_1,\wbb_1)\triangleq&\ \mathbb{E}_{\nb}\big[f_{\wbb_1}(\nb)]-\mathbb{E}_{\rbb}[f_{\wbb_1}(g_{\thetab_1}(\rbb))\big]
\label{eq:loss1} \\
\mathcal{L}_{\Zb}(\thetab_1,\thetab_2,\wbb_2)\triangleq&\ \mathbb{E}_{\Zb}\big[f_{\wbb_2}(\Zb)\big]-\ \mathbb{E}_{\Zb,\rbb}\big[f_{\wbb_2}(g_{\thetab_2}(\Zb)+g_{\thetab_1}(\rbb))\big]
\label{eq:loss2} \\
\mathcal{L}_{\text{cyc}}(\thetab_2,\thetab_3)\triangleq&\ \mathbb{E}_{\Zb}\big[\|\bm z-g_{\thetab_3}(g_{\thetab_2}(\Zb))\|_1\big].\label{eq:loss3}
\end{align}
The loss (\ref{eq:loss1}) is a standard W-GAN \citep{arjovsky2017wasserstein} loss for training the first generator-critic pair, $(g_{\thetab_1},f_{\wbb_1})$, of which $g_{\thetab_1}$ learns to generate the independent realization of the noise mimicking the patches in $\mathcal{N}=\{\nb^{(j)}\}_{j=1}^N$, taking the random vector $\rbb\sim \mathcal{N}(0,I)$ as input. The second loss (\ref{eq:loss2}) links the two generators, $g_{\thetab_1}$ and $g_{\thetab_2}$, with the second critic, $f_{\wbb_2}$. 
The second generator $g_{\thetab_2}$ is 
intended to generate the \textit{estimate} of the underlying clean patch for $\Zb$, \textit{i.e.}, coarsely denoise $\Zb$, and the critic $f_{\wbb_2}$ determines how close the distribution of the \textit{generated} noisy image, $g_{\thetab_2}(\Zb)+g_{\thetab_1}(\rbb)$, is to the that of $\Zb$\footnote{We assume $g_{\thetab_2}$ implicitly has the cropping step for $\Zb$ such that the dimension of $g_{\thetab_2}(\Zb)$ and $g_{\thetab_1}(\bm r)$ match.}. Our intuition is,
if $g_{\thetab_1}$ can realistically simulate the noise, then enforcing $g_{\thetab_2}(\Zb)+g_{\thetab_1}(\rbb)$ to mimick $\Zb$ would result in learning a reasonable initial \textit{denoiser} $g_{\thetab_2}$. One important detail regarding $g_{\thetab_2}$ is its final activation \textit{must} be the sigmoid function for stable training. The third loss (\ref{eq:loss3}), which resembles the cycle loss in \citep{zhu2017cyclegan}, imposes the encoder-decoder structure between $g_{\thetab_2}$ and $g_{\thetab_3}$, hence, helps $g_{\thetab_2}$ to compress the most redundant part of $\Zb$, \textit{i.e.}, the noise, and carry out the initial denoising. 
Once the losses are defined, training the generators and critics are done in an alternating manner, as in the training of W-GAN \citep{arjovsky2017wasserstein}, to approximately solve 
\begin{align}
\underset{\thetab_1,\thetab_2,\thetab_3}\min \  \underset{\wbb_1,\wbb_2}\max\Big[\ \alpha\Lcal_{\nb}(\thetab_1,\wbb_1)+&\ \beta\Lcal_{\Zb}(\thetab_1,\thetab_2,\wbb_2)+ \gamma\Lcal_{\text{cyc}}(\thetab_2,\thetab_3)\ \Big],\label{eq:overall_loss}
\end{align}
in which $(\alpha,\beta,\gamma)$ are hyperparameters to control the trade-offs between the loss functions.
\textcolor{black}{The pseudo algorithm for training a generative model is given in Algorithm \ref{alg:generative}.
There are a couple of subtle points for training with the overall objective (\ref{eq:overall_loss}), and we describe the full details on model architectures and hyperparameters 
in the S.M.}

\begin{algorithm}
        \caption{Training a generative model, all experiments in this paper used the defaults values, $n_{critic} = 5$, $n_{epoch} = 30$, m = 64, $\alpha_{g} = 4e^{-4}$, $\alpha_{critic} = 5e^{-5}$, $\alpha = 5$, $\beta = 1$, $\gamma = 10$}\label{alg:generative}
        \begin{algorithmic}[1]
            \STATE \textbf{Require} $\mathcal{D}$, $\lambda$
            \STATE $\mathcal{N}\gets$ NoisePatchExtraction$(\mathcal{D}, \lambda)$
            \FOR{$ep_{GAN}\gets 1, n_{epoch}$}
                \STATE Sample $\{n^{(i)}\}^{m}_{i=1} \sim \mathcal{N},\, \{r^{(i)}\}^{m}_{i=1} \sim N(0,I),\, \{Z^{(i)}\}^{m}_{i=1} \sim \mathcal{D}$
                \FOR{$ep_{critic}\gets 1, n_{critic}$}
                \STATE $g_{w_{1}}\gets \nabla_{w_{1}}[\mathcal{L}_{n}(\theta_{1},w_{1})]$, \  $g_{w_{2}}\gets \nabla_{w_{2}}[\mathcal{L}_{Z}(\theta_{1},\theta_{2},w_{2})]$
                \STATE $w_{1}\gets Clip(w_{1} + \alpha_{critic} \cdot Adam(w_{1}, g_{w_{1}}), -c, c)$
                \STATE $w_{2}\gets Clip(w_{2} + \alpha_{critic} \cdot Adam(w_{2}, g_{w_{2}}), -c, c)$
                \ENDFOR
            \STATE $g_{\theta_{1}}, g_{\theta_{2}}, g_{\theta_{3}} \gets \nabla_{\theta_{1}, \theta_{2}, \theta_{3}}[\ \alpha\Lcal_{n}(\theta_1,w_1)+\ \beta\Lcal_{Z}(\theta_1,\theta_2,w_2)+ \gamma\Lcal_{\text{cyc}}(\theta_2,\theta_3)\ ]$
            \STATE $\theta_{1}\gets \theta_{1} -\alpha_{g} \cdot Adam(\theta_{1}, g_{\theta_{1}})$
            , \ $\theta_{2}\gets \theta_{2} -\alpha_{g} \cdot Adam(\theta_{2}, g_{\theta_{2}})$
            , \ $\theta_{3}\gets \theta_{3} -\alpha_{g} \cdot Adam(\theta_{3}, g_{\theta_{3}})$
            \ENDFOR
            \STATE \textbf{return} $\theta_1, \theta_2$
        \end{algorithmic}
    \end{algorithm}

\vspace{-.05in}
\subsection{Iterative GAN2GAN training of a denoiser}\label{subsec:iterative gan2gan}

With our generative model, we then carry out the iterative Noisy N2N training as described in Section \ref{sec:motivation}, with the \textit{generated} noisy images. Namely, 
given each $\Zb^{(i)}\in\Dcal$, we generate the pair
\begin{equation}
(\hat{\Zb}_{11}^{(i)},\hat{\Zb}_{12}^{(i)})\triangleq(g_{\thetab_2}(\Zb^{(i)})+g_{\thetab_1}(\rbb^{(i)}_{11}), g_{\thetab_2}(\Zb^{(i)})+g_{\thetab_1}(\rbb^{(i)}_{12})),\label{eq:noisy pairs}
\end{equation}
in which $\rbb^{(i)}_{11},\rbb^{(i)}_{12}\in\mathbb{R}^{128}$ are i.i.d. $\sim\mathcal{N}(\boldsymbol{0},I)$.
In contrast to the ideal case in Section \ref{sec:motivation}, each generated image in (\ref{eq:noisy pairs}) is a noise-corrupted version of $g_{\thetab_2}(\Zb^{(i)})$,
in which the corruption is done by the \textit{simulated} noise $g_{\thetab_1}(\rbb)$. 
Denoting the set of such pairs as $\hat{\Dcal}_1=\{(\hat{\Zb}_{11}^{(i)},\hat{\Zb}_{12}^{(i)})\}_{i=1}^n$, a denoiser $\hat{X}_{\bm\phi}(\Zb)$ is trained by minimizing 
$
\mathcal{L}_{\text{G2G}}(\bm\phi,\hat{\Dcal}_1)\triangleq\frac{1}{n}\sum_{i=1}^n(\hat{\Zb}^{(i)}_{11}-\hat{X}_{\bm\phi}(\hat{\Zb}^{(i)}_{12}))^2.\label{eq:n2n_loss}
$
In $\mathcal{L}_{\text{G2G}}(\cdot)$, we only use the generated noisy images and do \emph{not} use the actual observed $\Zb^{(i)}$, hence, we dubbed our training as GAN2GAN (G2G) training. Now, denoting the learned denoiser as G2G$_1$ (with parameter $\bm \phi_1$), we can iterate the G2G training. For the $j$-th iteration (with $j\geq 2$), we generate 
\begin{equation}
\big(\hat{\Zb}_{j1}^{(i)},\hat{\Zb}_{j2}^{(i)}\big)\triangleq\big(\hat{X}_{\bm\phi_{j-1}}(\Zb^{(i)})+g_{\thetab_1}(\rbb^{(i)}_{j1}), \hat{X}_{\bm\phi_{j-1}}(\Zb^{(i)})+g_{\thetab_1}(\rbb^{(i)}_{j2})\big),\label{eq:noisy pairs_j}
\end{equation}
for each $\Zb^{(i)}$ and denote the resulting set of the pairs as $\hat{\mathcal{D}}_j$. Note in (\ref{eq:noisy pairs_j}), we \textit{update} the noisy estimate of the clean image with the output of 
G2G$_{j-1}$. Then, the new denoiser G2G$_j$ is obtained by computing
$\bm\phi_j \triangleq \arg\min_{\bm\phi}\mathcal{L}_{\text{G2G}}(\bm\phi,\hat{\mathcal{D}}_j),
$
 where the minimization is done via warm-starting from $\bm\phi_{j-1}$.
In our experiments, we show the sequence, G2G$_{j\geq1}$, successively refines the denoising quality and significantly improves the initial noisy estimate, similarly as in  Figure \ref{fig:noisy_n2n}. Moreover, we identify the benefit of the iterative G2G training becomes greater when noise is more sophisticated; \textit{i.e.}, for synthetic noise, the performance of G2G$_{j\geq1}$ converges after 1$\sim$3 iterations, whereas for the real-world microscopy noise, the performance keeps increasing until larger number of iterations.

\pdfoutput=1
\vspace{-.05in}
\section{Experimental results}
\vspace{-.02in}

\subsection{Data and experimental settings}\label{sec:data}
\vspace{-.02in}

\noindent\textbf{Data \& training details}\ \ In synthetic noise experiments, 
we always used the noisy training images from BSD400  \citep{MartinFTM01BSD}.
For evaluation, we used the standard BSD68 \citep{foe} as a test set. For real-noise experiment, we experimented on two data sets: the WF set in the microscopy image datasets in \citep{zhang2018poisson} and the reconstructed CT dataset. For both datasets, we trained/tested on each (Avg = $n$) and each dose level, respectively, which corresponds to different noise levels. 
For the generative model training, the patch size used for $\Dcal$ and $\mathcal{N}$ was  $96\times 96$, and $n$ and $N$ were set to $20,000$ (BSD) and $40,000$ (microscopy), respectively.  
For the iterative G2G training, the patch size for $\Dcal$ was $120\times120$ and $n=20,500$,
and in every mini-batch, we generated new noisy pairs with $g_{\thetab_1}$ as in the noise augmentation of \citep{ZhaZuoCheMenZha17}. The architecture of G2G$_j$ was set to 17-layer DnCNN in \citep{ZhaZuoCheMenZha17}. We put full details on training, model architectures and hyperparameters as well as the software platforms in the S.M. \vspace{.02in}\\
\noindent\textbf{Baselines}\ \ The baselines were BM3D \citep{bm3d}, DnCNN-B \citep{zhang2018residual}, N2N \citep{lehtinen2018noise2noise}, 
and N2V \citep{krull2018noise2void}. We reproduced and trained DnCNN-B, N2N and N2V using the publicly available source codes on the \emph{exactly} same training data as our iterative G2G training.
For DnCNN-B and N2N, which use either clean targets or two independent noisy image copies, we used 20-layers DnCNN model with composite additive white Gaussian noise with $\sigma\in[0, 55]$. 
N2V considers the same setting as ours and uses the \textit{exact} same architecture as  G2G$_j$; more details on N2V are also given in the S.M. 
We could not compare with the scheme in \citep{(HQDenoising)laine2019high}, since their code cannot run beyond white Gaussian noise case in our experiments and they had an unfair advantage: they \textit{newly} generate noisy images by corrupting given clean images for \textit{every} mini-batch whereas we assume the given noisy images are fixed once for all. It is known that such noise augmentation significantly can increase the performance, and their code could not run in our setting in which the noisy images are fixed once given. 
As an upper bound, we implemented N2C(Eq.(\ref{eq:dwt_smooth})), denoting a 17-layer DnCNN trained with clean target images in BSD400 and their noisy counterpart, which is corrupted by our $g_{\thetab_1}$ learned with  (\ref{eq:dwt_smooth}).

\subsection{Denoising results on synthetic noise}\label{subsec:synthetic}

\noindent\textbf{White Gaussian noise}\ \ Table \ref{table:bsd68_gaussian} 
shows the results on BSD68 corrupted by white Gaussian noise with different $\sigma$'s. Several variations of our G2G, $g_{\thetab_2}$ and the G2G iterates, G2G$_{j\geq 1}$, are shown for two different training data versions for learning the generative model.
\begin{table*}[h]\vspace{-.1in}\caption{Results on \emph{BSD68/Gaussian}. Boldface denotes algorithms that only use single noisy images. Red and blue denotes the highest and second highest result among those algorithms, respectively.}\label{table:medical50}
\centering
\smallskip\noindent
\resizebox{.98\linewidth}{!}{
\begin{tabular}{|c|c|c|c|c|c|c|c|c|c|}
\hline
\multirow{2}{*}{PSNR/SSIM} & \multicolumn{4}{c|}{Baselines}                                              & \multicolumn{4}{c|}{G2G variation}                                                            & Upper Bound  \\ \cline{2-10} 
                           & BM3D                  & DnCNN-B      & N2N          & N2V                   & $g_{\theta_{2}}$      & G2G$_1$               & G2G$_2$               & G2G$_3$               & N2C(Eq.(4))  \\ \hline
$\sigma = 15$              & \textbf{31.07/0.8717} & 31.44/0.8836 & 31.20/0.8745 & \textbf{29.48/0.8199} & \textbf{25.94/0.7519} & \textbf{30.98/0.8552} &\textbf{\color{red}{32.51}}/\textbf{\color{red}{0.8827}} & \textbf{\color{blue}{31.45}}/\textbf{\color{blue}{0.8825}} & 31.64/0.8870 \\
$\sigma = 25$              & \textbf{28.56/0.8013} & 28.92/0.8137 & 28.74/0.8041 & \textbf{26.97/0.7083} & \textbf{24.16/0.6630} & \textbf{28.23/0.7669} & \textbf{\color{blue}{28.82}}/\textbf{\color{blue}{0.8056}} & \textbf{\color{red}{28.96}}/\textbf{\color{red}{0.8080}} & 29.11/0.8189 \\
$\sigma = 30$              & \textbf{27.78/0.7727} & 28.06/0.7812 & 27.91/0.7720 & \textbf{26.38/0.6657} & \textbf{23.43/0.5967} & \textbf{27.58/0.7413} & \textbf{\color{blue}{27.99}}/\textbf{\color{red}{0.7783}} & \textbf{\color{red}{28.03}}/\textbf{\color{blue}{0.7759}} & 28.28/0.7890 \\
$\sigma = 50$              & \textbf{\color{blue}{25.60}}/\textbf{\color{red}{0.6866}} & 25.78/0.6721 & 25.71/0.6712 & \textbf{24.30/0.5765} & \textbf{20.58/0.4482} & \textbf{25.08/0.6215} & \textbf{25.55/0.6639} & \textbf{\color{red}{25.78}}/\textbf{\color{blue}{0.6749}} & 26.03/0.6951 \\ \hline
\end{tabular}}
    \vspace{-.1in}
\label{table:bsd68_gaussian}
\end{table*}
    Firstly, we clearly observe the iterative G2G training is \textit{very} effective; namely, it significantly improves the initial noisy estimate $g_{\thetab_2}$, particularly when the quality of the initial estimate is not good enough.
    This result confirms the result of Figure \ref{fig:noisy_n2n} indeed carries over to the blind denoising setting with our method. 
Secondly, we note G2G$_1$ already considerably outperforms
N2V, which is trained with the \textit{exact} same model architecture and dataset. 
Finally, the performance of G2G$_3$ is \textit{very} strong; it outperforms BM3D, which knows true $\sigma$, and even sometimes outperforms the blindly trained DnCNN-B and N2N, which is trained with the same BSD400 dataset, but with more information. This somewhat counter-intuitive result is possible since our G2G$_j$ accurately learns the correct noise level in the image, while DnCNN-B and N2N are trained with the composite noise levels, $\sigma\in[0,55]$.

\noindent\textbf{Mixture and correlated noise}\ \ Table \ref{table:bsd68_mixture_and_correlated} shows the results on mixture and correlated noise beyond white Gaussian. Note our G2G$_j$ does not assume any distributional or correlation structure of the noise, hence, it can still run as long as the assumption on the noise holds. 
\begin{table*}[h]\vspace{-.13in}
\caption{Results on \emph{BSD68/Mixture \& Correlated noise}. The boldface and colored texts are as before. 
}

\centering
\smallskip\noindent
\resizebox{.98\linewidth}{!}{
\begin{tabular}{|c|c|c|c|c|c|c|c|c|c|c|c|}
\hline
\multicolumn{3}{|c|}{\multirow{2}{*}{PSNR/SSIM}}                                                                   & \multicolumn{4}{c|}{Baselines}                                              & \multicolumn{4}{c|}{G2G variation}                                                            & Upper bound  \\ \cline{4-12} 
\multicolumn{3}{|c|}{}                                                                                             & BM3D                  & DnCNN-B      & N2N          & N2V                   & $g_{\theta_{2}}$      & G2G$_1$               & G2G$_2$               & G2G$_3$               & N2C(Eq.(4))  \\ \hline
\multirow{4}{*}{\begin{tabular}[c]{@{}c@{}}Mixture\\ noise\end{tabular}} & \multirow{2}{*}{Case A} & $s = 15$      & \textbf{41.44/0.9822} & 39.62/0.9749 & 40.59/0.9860 & \textbf{33.53/0.9368} & \textbf{31.85/0.9522} & \textbf{42.35/0.9876} & \textbf{\color{red}{42.56}}/\textbf{\color{red}{0.9888}} & \textbf{\color{blue}{42.49}}/\textbf{\color{blue}{0.9885}} & 42.92/0.9843 \\
                                                                         &                         & $s = 25$      & \textbf{37.97/0.9647} & 37.23/0.9616 & 37.39/0.9737 & \textbf{31.62/0.9057} & \textbf{32.73/0.9478} & \textbf{39.13/0.9761} & \textbf{\color{blue}{39.64}}/\textbf{\color{blue}{0.9800}} & \textbf{\color{red}{39.72}}/\textbf{\color{red}{0.9807}} & 40.42/0.9843 \\ \cline{2-12} 
                                                                         & \multirow{2}{*}{Case B} & $s = 30$      & \textbf{30.12}/\textbf{\color{red}{0.8549}} & 30.58/0.8655 & 30.36/0.8559 & \textbf{28.10/0.7543} & \textbf{27.55/0.7728} & \textbf{29.05/0.8199} & \textbf{\color{blue}{30.32}}/\textbf{0.8456} & \textbf{\color{red}{30.49}}/\textbf{\color{blue}{0.8538}} & 30.78/0.8685 \\
                                                                         &                         & $s = 50$      & \textbf{29.27/0.8190} & 30.20/0.8547 & 30.20/0.8547 & \textbf{28.22/0.7755} & \textbf{27.36/0.7712} & \textbf{29.78/0.8345} & \textbf{\color{red}{30.04}}/\textbf{\color{blue}{0.8392}} & \textbf{\color{blue}{30.00}}/\textbf{\color{red}{0.8417}} & 30.39/0.8574 \\ \hline
\multicolumn{2}{|c|}{\multirow{2}{*}{\begin{tabular}[c]{@{}c@{}}Correlated\\ noise\end{tabular}}}  & $\sigma = 15$ & \textbf{29.84/0.8504} & 30.84/0.9011 & 30.69/0.9223 & \textbf{28.80/0.8367} & \textbf{28.13/0.8370} & \textbf{30.73/0.8889} & \textbf{\color{blue}{31.09}}/\textbf{\color{blue}{0.8949}} & \textbf{\color{red}{31.26}}/\textbf{\color{red}{0.8954}} & 31.60/0.9075 \\
\multicolumn{2}{|c|}{}                                                                             & $\sigma = 25$ & \textbf{26.69/0.7544} & 27.39/0.8257 & 27.32/0.8594 & \textbf{26.11/0.7348} & \textbf{25.68/0.7607} & \textbf{27.80/0.8130} & \textbf{\color{red}{28.01}}/\textbf{\color{blue}{0.8271}} & \textbf{\color{blue}{28.00}}/\textbf{\color{red}{0.8447}}             & 28.42/0.8376 \\ \hline
\end{tabular}}
    \label{table:bsd68_mixture_and_correlated}
        \vspace{-.08in}
\end{table*}
In the table,  the G2G results are for (BSD) as specified above. Moreover, DnCNN-B and N2N are still blindly trained with the \textit{mismatched} white Gaussian noise. For mixture noise, we tested with two cases. Case A corresponds to the same setting as given in \citep{chen2018image}, \textit{i.e.}, 70\% $\sim\mathcal{N}(0,0.1^2)$, 20\% $\sim\mathcal{N}(0,1)$, and 10\% $\sim$ Unif$[-s,s]$ \textcolor{black}{which means the random variable that is uniformly distributed between $[-s,s]$} with $s=15, 25$. For case B, we tested with larger variances, \textit{i.e.}, 70\% Gaussian $N(0,15^2)$, 20\% Gaussian $N(0,25^2)$, and 10\% Uniform $[-s,s]$ with $s=30, 50$. 
\textcolor{black}{For correlated noise, we generated the following noise for each $\ell$-th pixel,
\begin{equation}
    N_\ell = \eta M_\ell + (1-\eta) \Big(\frac{1}{\sqrt{|\mathcal{NB}_{\ell}|}}\sum_{m\in\mathcal{NB}_{\ell}}M_m\Big), \ \ \ell=1,2, \ldots\nonumber
\end{equation}
 in which $\{M_{\ell}\}$ are white Gaussian $\mathcal{N}(0,\sigma^2)$, $\mathcal{NB}_{\ell}$ is the $k\times k$ neighborhood patch except for the pixel $\ell$, and $\eta$ is a mixture parameter. We set $\eta=1/\sqrt{2}$ such that the marginal distribution of $N_\ell$ is also $\mathcal{N}(0,\sigma^2)$ and set $k=16$. Note in this case, $N_{\ell}$ has a spatial correlation, and we tested with $\sigma=15, 25$. }
 From the table, we first note that 
DnCNN-B and N2N suffer from serious performance degradation for both mixture and correlated noises due to noise mismatch, and the conventional BM3D outperforms them for some cases (\textit{e.g.}, Case A for mixture noise). However, we note our G2G$_2$ can still denoise very well after just two iterations and outperforms all the baselines for all noise types. Note N2V seriously suffers and is \textit{not} comparable to ours. Finally, N2C(Eq.(\ref{eq:dwt_smooth})) is a sound upper bound for all noise types, confirming the correctness of the extraction rule (\ref{eq:dwt_smooth}).

\subsection{Denoising results on real noise}

We also test our method on the real-world noise. While some popular real noise is known to have source-dependent characteristics, there are also cases in which the noise is source-independent and pixel-wise correlated, which satisfies the assumption of our method. We tested on two such datasets, the Wide-Focal (WF) set in the microscopy image dataset \citep{zhang2018poisson} and a Reconstructed CT dataset.
A more detailed description and analysis on these two datasets are in S.M.
 \begin{figure}[h]\vspace{-.1in}
\centering  
\subfigure[Wide-Focal (WF)]
{\includegraphics[width=0.33\linewidth]{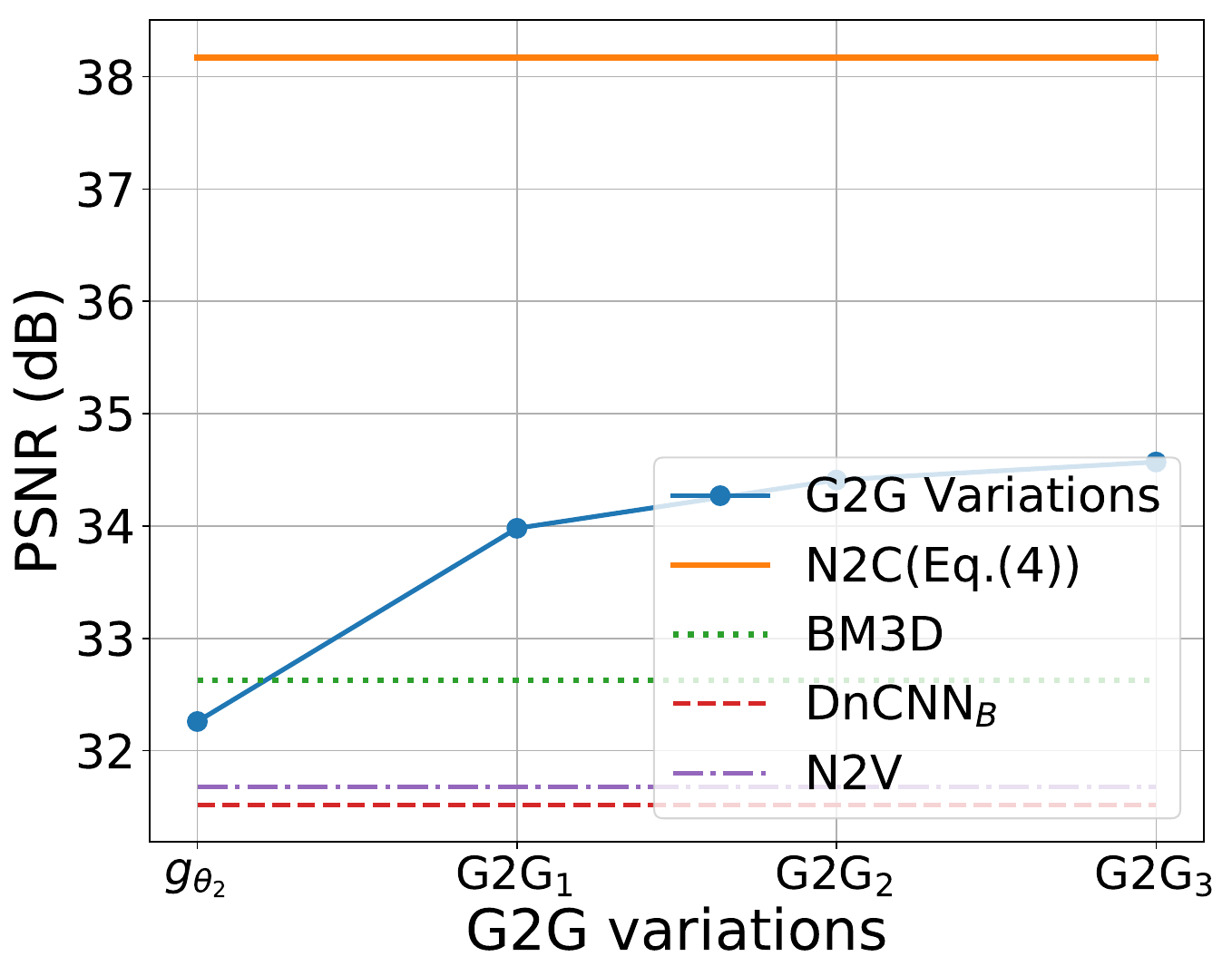}
\label{figure:real_noise_psnr}}
\subfigure[Reconstructed CT]
{\includegraphics[width=0.33\linewidth]{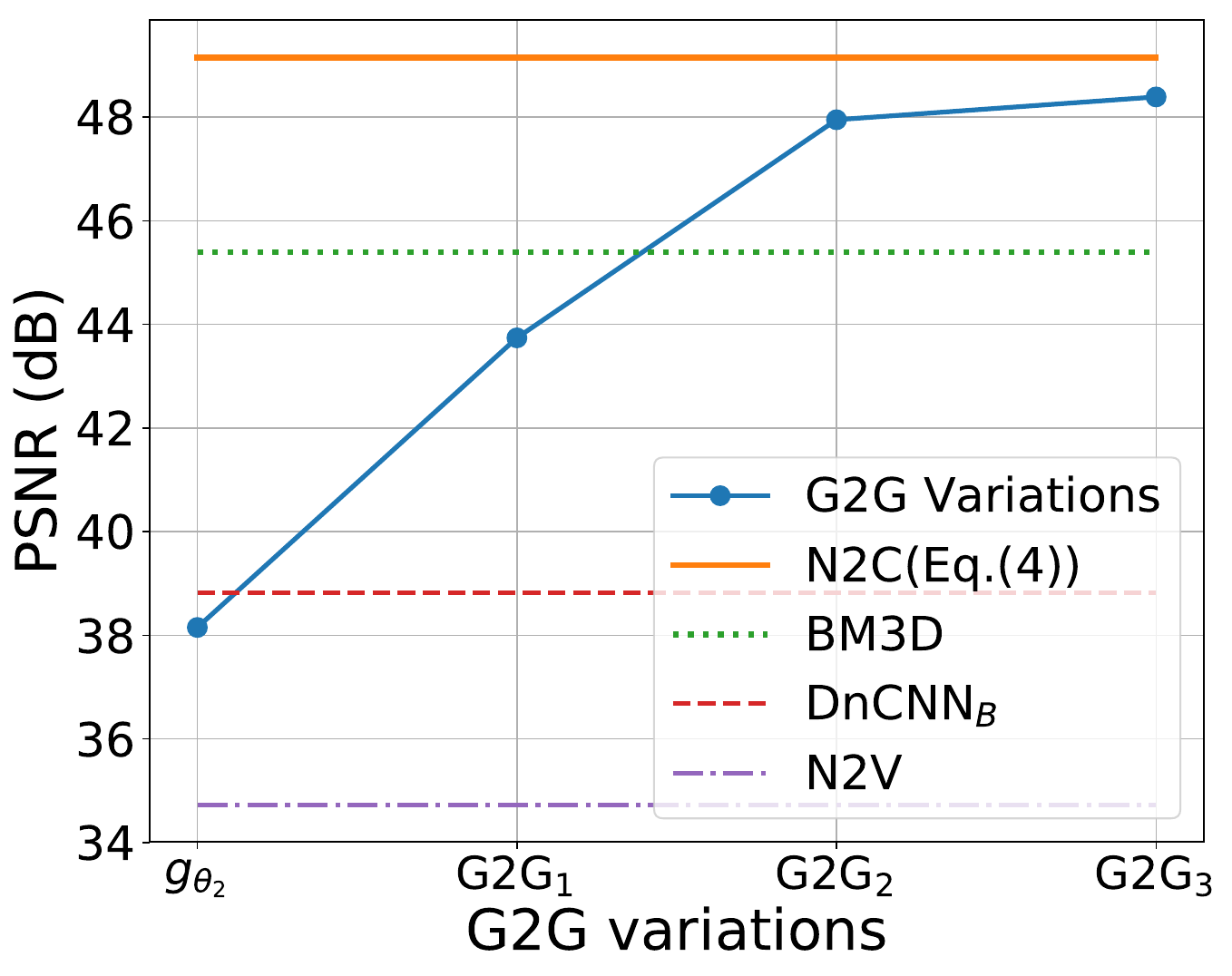}
\label{figure:real_noise_ssim}}\hspace{.02in}
\subfigure[Visualization]
{\includegraphics[width=0.23\linewidth]{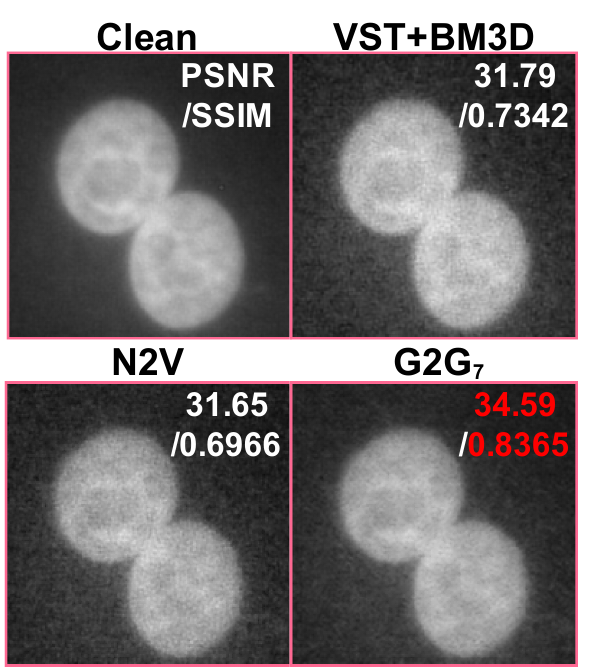}
\label{figure:visualization}}\vspace{-.13in}
\caption{Results on real microscopy image denoising on WF and medical image denoising.}\vspace{-.1in}\label{figure:real_noise}
 \end{figure}
The WF and Reconstructed CT data
has 5 sets (Avg = $1,2,4,8,16$) and 4 sets (Dose=$25,50,75,100$) with different noise levels, respectively. We did \emph{not} exploit the fact that the images are multiple noisy measurements of a clean image, which enables employing N2N, but treated them as noisy images of distinct clean images. 
Figure \ref{figure:real_noise_psnr} and \ref{figure:real_noise_ssim} shows the PSNR of all methods for each dataset, respectively, averaged over all sets.
The baselines were DnCNN-B, BM3D 
and N2V. We note BM3D estimated noise $\sigma$ using the method in \cite{(NoiseEst)chen2015efficient}.
We iterated until G2G$_3$ and N2C(Eq.(\ref{eq:dwt_smooth})) was an upper bound for each set. 
We clearly observe that the performance of G2G$_j$ significantly improves (over $g_{\bm\theta_2}$) as the iteration continues.
In results,  G2G$_{3}$ becomes significantly better than DnCNN-B and N2V as well as BM3D, still one of the strongest baselines for real-world noise denoising when no clean target images are available, for both datasets. 
We report more detailed experimental results (including SSIM) on both datasets in S.M.
Moreover, the inference time for BM3D is about 4.5$\sim$5.0 seconds per image since a noise estimation has to be done for each image separately, whereas that for G2G$_j$ is only 4 ms (on GPU), which is another significant advantage of our method. 
Figure \ref{figure:visualization} shows the visualizations on the WF, and we give more examples in the S.M.

\vspace{-.05in}
\subsection{Ablation study}\label{sec:ablation}
\vspace{-.05in}

\begin{wrapfigure}{r}{0.5\textwidth}
\vspace{-.05in}
    \centering
    \includegraphics[width=0.5\textwidth]{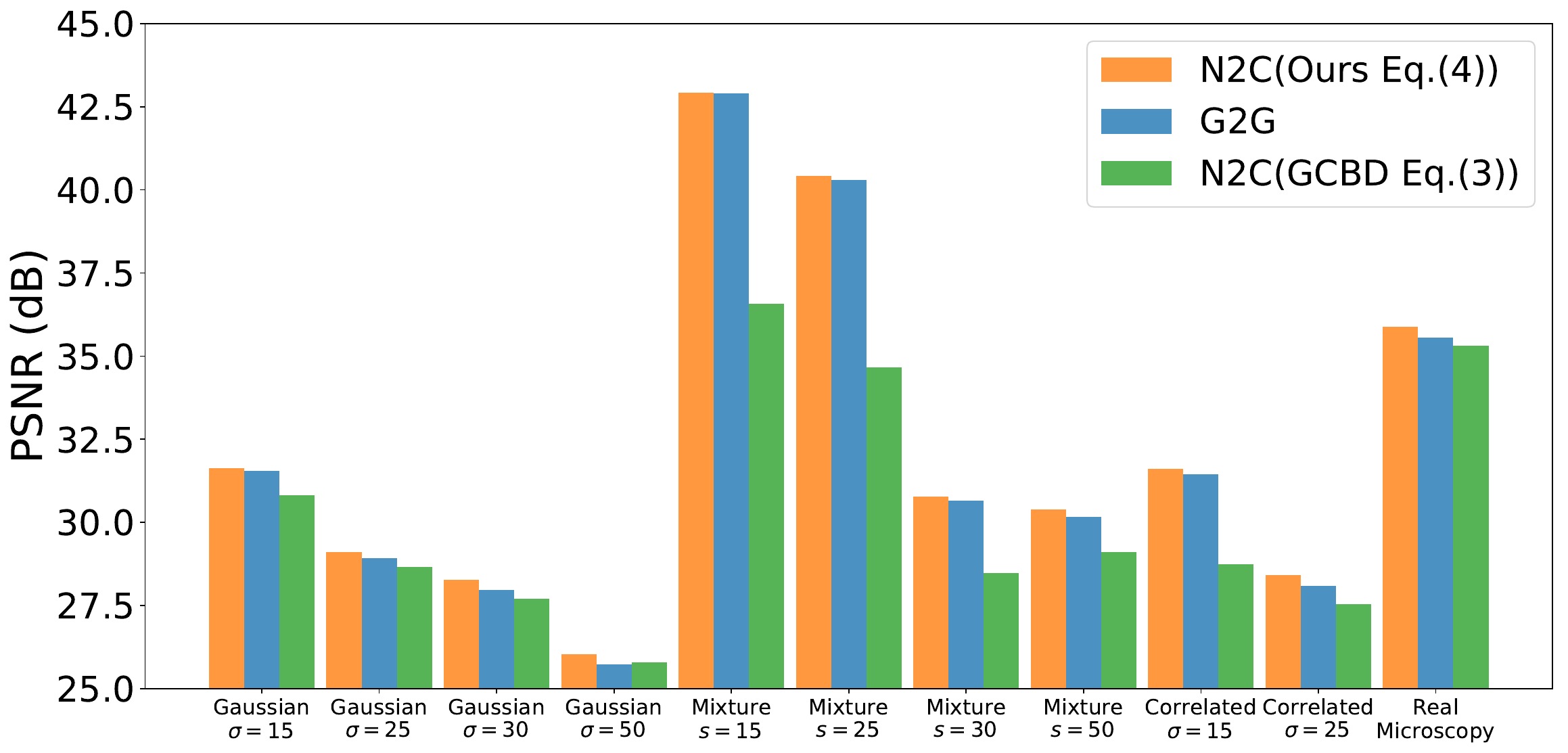}
    \vspace{-.12in}
    \caption{Effect of noise patch extraction rule.
    }\label{fig:gcbd_compare_models}
    \vspace{-.13in}
\end{wrapfigure}

\noindent\textbf{Noise patch extraction} \ 
Here, we evaluate the effect of the noisy patch extraction rules (\ref{eq:var_smooth}) and (\ref{eq:dwt_smooth}) in the final denoising performance. Figure \ref{fig:gcbd_compare_models} compares the PSNR of N2C(GCBD Eq.(\ref{eq:var_smooth})), a re-implementation of \citep{chen2018image}, N2C(Ours Eq.(\ref{eq:dwt_smooth})) and the best G2G, for each dataset. 
We note neither source code nor training data of \citep{chen2018image} is publicly available, and the PSNR in \citep{chen2018image} could not be reproduced (with the exact same $\eta$ and $\gamma$ as in \citep{chen2018image}). 
From the figure, we clearly observe the significant gap between N2C(Our Eq.(\ref{eq:dwt_smooth})) and N2C(GCBD Eq.(\ref{eq:var_smooth})), particularly when the noise is not white Gaussian. Moreover, our \textit{pure} unsupervised G2G with (\ref{eq:dwt_smooth}) even outperforms N2C(GCBD Eq.(\ref{eq:var_smooth})) that utilizes the clean target images, confirming the quality difference shown 
in Figure \ref{figure:patches} significantly affects learning noise and a denoiser.

\begin{figure}[h]\vspace{-.1in}
\centering  
\subfigure[Variations of $g_{\thetab_2}$]
{\includegraphics[width=0.32\linewidth]{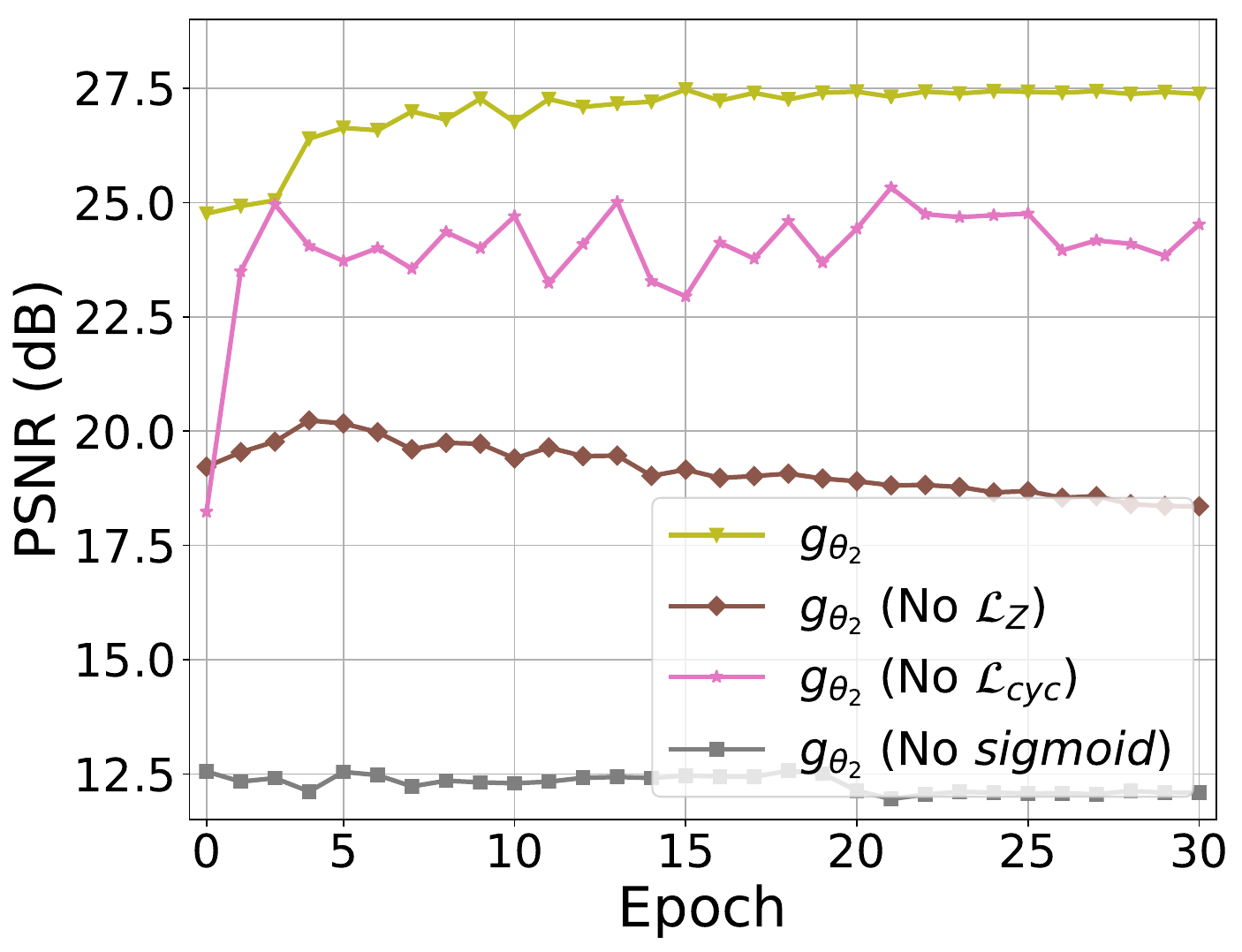}
\label{figure:ablation_gan}}
\subfigure[Iterative G2G (Synthetic)]
{\includegraphics[width=0.32\linewidth]{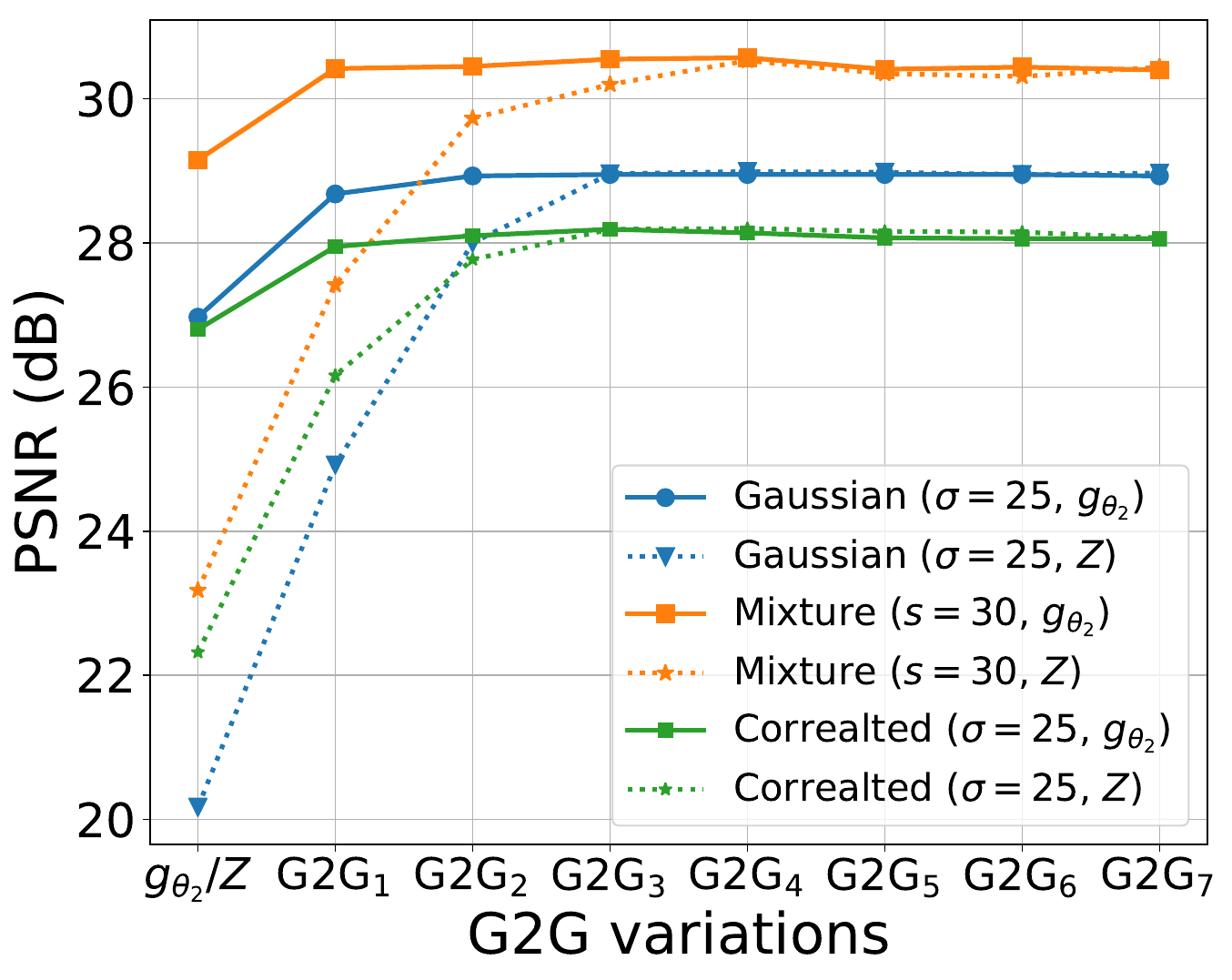}
\label{figure:ablation_iterative_synthetic}}
\subfigure[Iterative G2G (Real microscopy)]
{\includegraphics[width=0.32\linewidth]{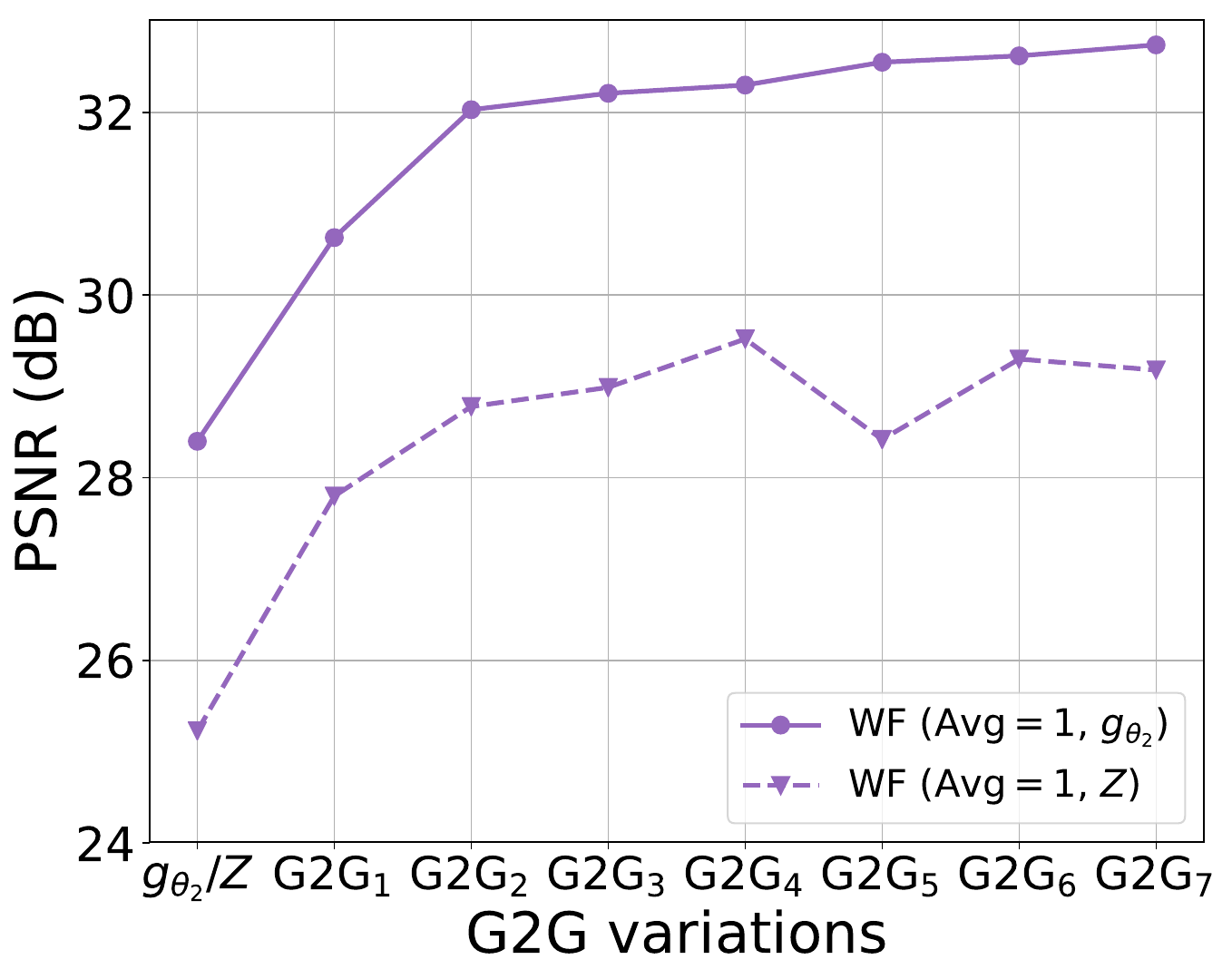}
\label{figure:ablation_iterative_real}}\vspace{-.1in}
\caption{Ablation studies. (b) and (c) compare the performances between starting from $g_{\thetab_2}$ and $\Zb$.}\vspace{-.03in}

 \end{figure}

\noindent\textbf{Generative model and iterative G2G training}\ \
 Figure \ref{figure:ablation_gan} 
 shows the PSNRs of $g_{\thetab_2}$
  on BSD68/Gaussian($\sigma=25$) trained with 
 three variations; ``No $\mathcal{L}_{\Zb}$'' for no $f_{\wb_2}$, ``No $\mathcal{L}_{\text{cyc}}$'' for no (\ref{eq:loss3}) and $g_{\thetab_3}$, and ``No sigmoid'' for no sigmoid activation at the output layer of $g_{\thetab_2}$. 
 We confirm that our proposed architecture achieves the highest PSNR for  $g_{\thetab_2}$, the sigmoid activation and $f_{\wb_2}$ are essential, and the cylce loss (\ref{eq:loss3}) is also important. 
 Achieving a decent PSNR for 
 $g_{\thetab_2}$ is beneficial for saving the number of G2G iterations and achieving high final PSNR. More detailed analyses on the generative model architecture are in the S.M.
 Figure \ref{figure:ablation_iterative_synthetic} and \ref{figure:ablation_iterative_real} show the effect of the quality of the initial estimate for the iterative G2G training. From Figure \ref{fig:noisy_n2n}, one may ask whether $g_{\thetab_2}$ is indeed necessary, since even when $\sigma_0\approx\sigma$, the iterating the Noisy N2N can mostly achieve the upper bound. Hence, for samples of synthetic and real microscopy data, we evaluate how G2G$_j$ performs when the iteration simply starts with $\Zb$. Figure \ref{figure:ablation_iterative_synthetic} shows a somewhat surprising result that for synthetic noises, 
 starting from $\Zb$ achieves essentially the same performance as starting from $g_{\thetab_2}$ with a couple more G2G iterations. However, for real microscopy noise case in Figure \ref{figure:ablation_iterative_real}, WF(Avg$=1$) in which starting from $\Zb$ achieves far lower performance than starting from $g_{\thetab_2}$, justifying our generative model for attaining the initial noisy estimate.

\vspace{-.05in}

\section{Concluding Remark}

\textcolor{black}{Motivated by a novel observation on Noisy N2N, we proposed a novel GAN2GAN method, which can tackle the challenging blind image denoising problem solely with single noisy images. Our method showed impressive denoising performance that even sometimes outperform the methods with more information as well as VST+BM3D for real noise denoising. As a future work, we plan to extend our framework to more explicitly handle the source-dependent real-world noise.}

\section*{Acknowledgment}

This work was supported in part by NRF Mid-Career Research Program [NRF-2021R1A2C2007884] and IITP grant [No.2019-
0-01396, Development of framework for analyzing, detecting, mitigating of bias in AI model and
training data], funded by the Korean government (MSIT).





\newpage

\newpage

\bibliographystyle{iclr2021_conference}
\bibliography{bibfile.bib}

\end{document}


\maketitle

\section{Proof of Theorem 1}
\begin{theorem}\label{thm:mse}
Consider the single-letter Gaussian setting and 
$f_{\text{Noisy N2N}}(Z,y)$ obtained in (Eq.(2), manuscript). Also, assume  $0<y=\sigma_0^2/\sigma_X^2<1$. Then, there exists some $y_0$ s.t. $\forall y\in(y_0,1)$, $\mathbb{E}(X-f_{\text{Noisy N2N}}(Z,y))^2<\sigma_0^2$. 
\end{theorem}
\textit{Proof:} We consider the following chain of equalities:
\begin{eqnarray}
&&\sigma_0^2-\mathbb{E}(X-f_{\text{Noisy N2N}}(Z,y))^2\\
&=&\sigma_0^2-\mathbb{E}\Big(X-\frac{\sigma_X^2(1+y)}{\sigma_X^2(1+y)+\sigma_N^2}(X+N)\Big)^2\\
&=& \sigma_0^2-\mathbb{E}\Big(\frac{\sigma_N^2}{\sigma_X^2(1+y)+\sigma_N^2}X-\frac{\sigma_X^2(1+y)}{\sigma_X^2(1+y)+\sigma_N^2}N\Big)^2\\
&=&\sigma_X^2\Big(y-\frac{\sigma_N^4}{(\sigma_X^2(1+y)+\sigma_N^2)^2}-\frac{\sigma_N^2\sigma_X^2(1+y)^2}{(\sigma_X^2(1+y)+\sigma_N^2)^2}\Big)\\
&=&\sigma_X^2 \frac{y(\sigma_X^2(1+y)+\sigma_N^2)^2-\sigma_N^4-\sigma_N^2\sigma_X^2(1+y)^2}{(\sigma_X^2(1+y)+\sigma_N^2)^2}\label{eq:last}
\end{eqnarray}
Now, by denoting the numerator of (\ref{eq:last}) as $g(y)$, we have
\begin{eqnarray}
g(y)&=& y \big(\sigma_X^4(1+y)^2+\sigma_N^4 +2\sigma_X^2\sigma_N^2(1+y)\big)-\sigma_N^2\sigma_X^2(y^2+2y+1)-\sigma_N^4\\
&=&\sigma_X^4y^3 +(2\sigma_X^4+\sigma_N^2\sigma_X^2)y^2 +(\sigma_X^4+\sigma_N^4)y-\sigma_N^2\sigma_X^2-\sigma_N^4.
\end{eqnarray}
Then, we can easily see that
\begin{eqnarray}
g(0)&=&-\sigma_N^2\sigma_X^2-\sigma_N^4 <0\\
g(1)&=&4\sigma_X^4>0\\
g(y)&=& 3\sigma_X^4y^2+2(2\sigma_X^4+\sigma_N^2\sigma_X^2)y+\sigma_X^4+\sigma_N^4>0.
\end{eqnarray}
Therefore, in $0< y <1$, we can see $g(y)$ is an increasing function and has a root $y_0$ in the interval. Hence, the claim of the theorem: for all $y\in (y_0,1)$, $\mathbb{E}(X-f_{\text{Noisy N2N}}(Z,y))^2<\sigma_0^2$ holds. \qed

\section{Details on the experimental settings}
\subsection{Smooth noisy patch extraction}

\subsubsection{GCBD \citep{chen2018image} rule Eq.(3)}

The original GCBD paper \citep{chen2018image} did not provide any source code or training data, hence, we reproduced their noisy patch extraction algorithm. 
There are six hyperparameters for the rule [Eq.(3), Manuscript], and we used the exact same hyperparameters given in their paper, which are shown in Table \ref{table:gcbd_hyperparameter}.
$d$ and $h$ denote the size of a patch, $\bm p$, and its sub-patches, $\mathbf{q}_{j}$, given in  [Eq.(3), Manuscript], respectively. $s_{p}$ and $s_{q}$ are the stride sizes for extracting the patches, $\bm p$ and $\{\mathbf{q}_{j}\}$, from a given image. $\mu$ and $\lambda$ are the hyperparameters of the rule for selecting the smooth patches shown in [Eq.(3), Manuscript].

\begin{table}[h]\caption{Hyperparameters for the patch extraction rule of GCBD}

\centering
\smallskip\noindent
\resizebox{.6\linewidth}{!}{
\begin{tabular}{|c||c|c|c|c|c|c|}
\hline
Hyperparameters & $d$ & $h$ & $s_{p}$ & $s_{q}$ & $\mu$ & $\gamma$ \\ \hline
Values           & 64  & 16  & 32      & 16      & 0.1 & 0.25     \\ \hline
\end{tabular}}
    \label{table:gcbd_hyperparameter}
\end{table}

\subsubsection{G2G rule Eq.(4)}

There are three hyperparemeters for our extraction rule [Eq.(4 ), Manuscript], $\lambda$, $d$ (the patch size), and $s_d$ (the stride size for extracting patches from an image). The choices for our experiments are shown in Table \ref{table:lambda}. Moreover, we stress that we did \textit{not} 
tune $\lambda$ using clean images, but the different $\lambda$ values in the table are determined by the pre-determined number of extracted patches by applying our rule [Eq.(4), Manuscript]. Moreover, as argued in Section 3.1 (manuscript), we do not require any sub-patches to be extracted, hence, have only half the hyperparameters compared to the GCBD rule. 


\begin{table}[h]\caption{Hyperparameters for the extraction rule of G2G}

\centering
\smallskip\noindent
\resizebox{.5\linewidth}{!}{
\begin{tabular}{|c||c|c|c||c|c|}
\hline
                         & \begin{tabular}[c]{@{}c@{}}Gaussian\\ Noise\end{tabular} & \begin{tabular}[c]{@{}c@{}}Mixture\\ Noise\end{tabular} & \begin{tabular}[c]{@{}c@{}}Correlated\\ Noise\end{tabular} & WF   & Medical    \\ \hline
$\lambda$ & 0.03                                                     & 0.1                                                     & 0.15                                                       & 0.42 & 0.015 \\ \hline
$d$                      & \multicolumn{5}{c|}{96}                                                                                                                                                                        \\ \hline
$s_{p}$                  & \multicolumn{5}{c|}{24}                                                                                                                                                                        \\ \hline

\end{tabular}}
    \label{table:lambda}
\end{table}

\subsubsection{Effect of $\lambda$}
Table \ref{table:lambda} shows the effect of $\lambda$ in [Eq.(4), Manuscript] on the final performance of G2G$_2$. Note the smaller the $\lambda$, the less number of patches are extracted, but the homogeneity increases. The table shows $\lambda$ clearly affects the denoising performance of $g_{\bm\theta_2}$, but as the iterative G2G training continues, the performance of G2G$_{2}$ becomes not very sensitive to $\lambda$. Hence, in our experiments, we did not optimize $\lambda$ based on \emph{any} clean validation set, but just set $\lambda$ based on the number of extracted patches and checking the visual qualities of the patches. 

\begin{table*}[h]\caption{Effects of varying $\lambda$ on the denoising performance.}
\centering
\smallskip\noindent
\resizebox{.99\linewidth}{!}{
\begin{tabular}{|c|c|c|c||c|c|c|c||c|c|c|c|}
\hline
\multicolumn{4}{|c||}{Gaussian Noise ($\sigma = 25$)}        & \multicolumn{4}{c||}{Mixture Noise ($s = 25$)}               & \multicolumn{4}{c|}{Correlated Gaussian Noise ($\sigma = 25$)} \\ \hline
$\lambda$ & \# of patches & g$_{\theta_{2}}$ & G2G$_{2}$    & $\lambda$ & \# of patches & g$_{\theta_{2}}$ & G2G$_{2}$    & $\lambda$  & \# of patches  & g$_{\theta_{2}}$  & G2G$_{2}$    \\ \hline
0.03      & 100,000       & 26.30/0.7123     & \textbf{28.93/0.8293} & 0.1       & 80,000        & 32.73/0.9478     & \textbf{40.30/0.9845} & 0.15       & 100,000        & 25.68/0.7606      & \textbf{27.85/0.8185} \\ \hline
0.01      & 61,000        & 27.20/0.7159     & \textbf{28.84/0.8045} & 0.005     & 45,000        & 35.84/0.9588     & \textbf{40.16/0.9838} & 0.11       & 30,000         & 26.61/0.7566      & \textbf{27.67/0.8203} \\ \hline
0.0075    & 32,000        & 26.44/0.7085     & \textbf{28.80/0.8060} & 0.025     & 23,000        & 34.20/0.9398     & \textbf{40.28/0.9848} & 0.1        & 11,000         & 26.30/0.7440      & \textbf{27.65/0.8203} \\ \hline
\end{tabular}}
    \label{table:lambda}    
\end{table*}

\subsection{Training a W-GAN based generative model}\label{sec:wgan_based_generative_model}

Here, we elaborate a couple of subtle points for training our generative model as mentioned in Section 3.2 (manuscript).

\begin{figure}[h]\vspace{-.1in}
\centering  
\includegraphics[width=0.4\linewidth]{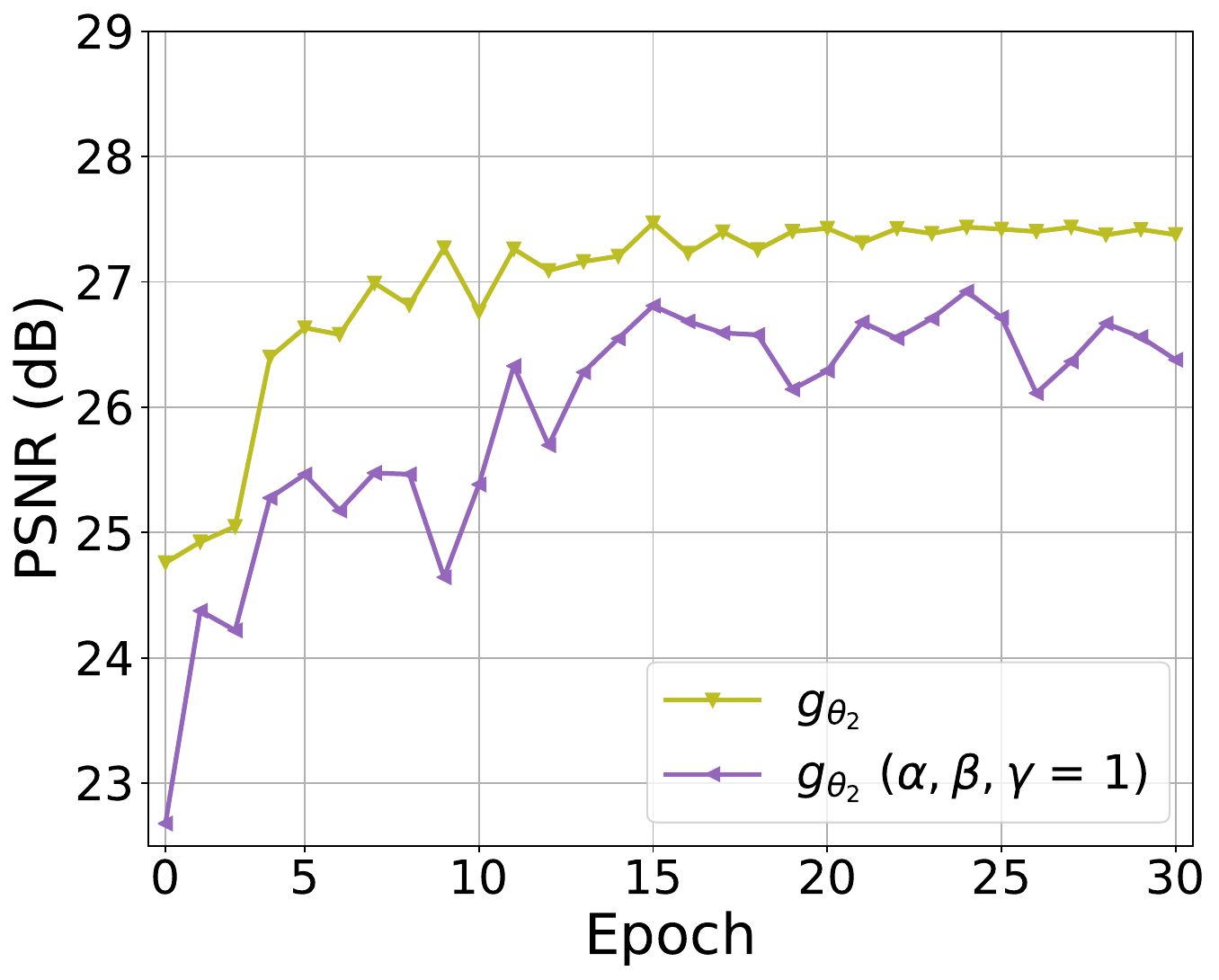}
\label{figure:ablation_gan}
\caption{Ablation study on $(\alpha,\beta,\gamma)$}
\end{figure}

Firstly, given the overall optimization objective [Eq.(8), manuscript], we use $(\alpha,\beta,\gamma)=(1,1,0)$ for the inner maximization for critics, and use $(\alpha,\beta,\gamma)=(5,1,10)$ for the outer minimization for generators. The main intuition for using different $(\alpha,\beta,\gamma)$ for training the generators is due to different levels of confidence in the generator loss terms. Namely, we assign the largest weight to [Eq.(7), Manuscript] since it is a deterministic loss and its value has a clear meaning. The generator loss [Eq.(5), Manuscript], which is in the form of the standard W-GAN loss, gets the medium level weight since the meaning of its value is less certain than [Eq.(6), Manuscript]. In contrast, the generator loss in [Eq.(5), Manuscript], which consists of two generators, can become somewhat unstable during training, hence, it gets the least weight. Figure \ref{figure:ablation_gan} compares the performance of $g_{\bm\theta_2}$'s on BSD68 $(\sigma = 25)$  when using $(\alpha,\beta,\gamma)=(5,1,10)$, for the outer minimization, as proposed, and using $(\alpha,\beta,\gamma)=(1,1,1)$. We observe there is a significant gap between the two. 

Secondly, the output layer of $g_{\thetab_2}$ \emph{must} have the sigmoid activation function. Note $g_{\thetab_2}$ itself can be thought of another denoiser, but since we are not training it with any target, we need to ensure the outputs of $g_{\thetab_2}$ have values between $[0,1]$ to prevent from obvious errors of generating negative or out-of-bound pixel values. Without the sigmoid activation, it turned out all the generators cannot be trained properly at all. 

Finally, using the right architectures for the generators and critics, \emph{e.g.}, number of layers and filters, was critical since the training procedure got very sensitive to the architectural variations. Tables \ref{table:details_generator} shows the details on the architecture of our first generator, $g_{\bm\theta_1}$, which aims to generate noise patches. 
The dimension of $\bm r$ (the input random vector) was set to 128, and $C$ denotes the channel of the generated noise patch. 
The architectures of the $g_{\bm\theta_2}$ and $g_{\bm\theta_3}$ in our generative model are equal to that of the DnCNN model \citep{zhang2018residual}, however, $g_{\thetab_2}$ had 15 layers with sigmoid activation in the output layer, and $g_{\thetab_3}$ had 17 layers and linear activation in the output layer. In addition, the architectures of the two critics,  $\{f_{\wbb_1}$,$f_{\wbb_2}\}$, in our generative model are given in Table \ref{table:details_discriminator}. 

\begin{table*}[h]\caption{Architectural details on $g_{\thetab_1}$.}
\centering
\smallskip\noindent
\resizebox{.8\linewidth}{!}{
\begin{tabular}{|c|c|c|c|c|c|c|}
\hline
\multicolumn{2}{|c|}{Input shape : $(128,)$}     & \multicolumn{5}{c|}{Details of DeConv layer}    \\ \hline
Layer Num      & Layer composition             & Input channel & Output channel & Kernel size & Stride & Padding \\ \hline
1              & DeConv + BatchNorm + ReLU     & 128   & 64     & 4           & 1      & 0       \\ \hline
2              & DeConv + BatchNorm + ReLU     & 64    & 32     & 4           & 2      & 1       \\ \hline
3              & DeConv + BatchNorm + ReLU     & 32    & 16     & 4           & 2      & 1       \\ \hline
4              & DeConv + BatchNorm + ReLU     & 16    & 8      & 4           & 1      & 1       \\ \hline
5              & Conv + Tanh                   & 8     & $C$      & 4           & 2      & 1       \\ \hline
\multicolumn{2}{|c|}{Output shape : (64x64x$C$)} & \multicolumn{5}{c|}{-}                          \\ \hline
\end{tabular}}
    \label{table:details_generator}
\end{table*}

\begin{table*}[h]\caption{Architectural details on the critics, $\{f_{\wbb_1}$, $f_{\wbb_2}\}$.}
\centering
\smallskip\noindent
\resizebox{.8\linewidth}{!}{
\begin{tabular}{|c|c|c|c|c|c|c|c|}
\hline
\multicolumn{2}{|c|}{Input shape : (64x64x$C$)}  & \multicolumn{5}{c|}{Details of Conv layer}      & Details of LeakyReLU \\ \hline
Layer Num     & Layer composition              & Input channel & Output channel & Kernel size & Stride & Padding & $\alpha$                \\ \hline
1             & Conv + BatchNorm + LeakyReLU   & C     & 128    & 4           & 2      & 1       & \multirow{3}{*}{0.2} \\ \cline{1-7}
2             & Conv + BatchNorm + LeakyReLU   & 128   & 256    & 4           & 2      & 1       &                      \\ \cline{1-7}
3             & Conv + BatchNorm + LeakyReLU   & 256   & 512    & 4           & 2      & 1       &                      \\ \hline
4             & Conv                           & 512   & 1      & 4           & 1      & 0       & -                    \\ \hline
\multicolumn{2}{|c|}{Output shape : (64x64x1)} & \multicolumn{5}{c|}{-}                          & -                    \\ \hline
\end{tabular}}
    \label{table:details_discriminator}
\end{table*}

For training, we carry out the random cropping of the given patches to the size of $64\times64$, and the data augmentation was done by flipping the cropped patches horizontally and vertically. For optimization, we used Adam \citep{KinBa15} optimizer for the three generators and RMSProp \citep{TieHin12} optimizer for the two critics. The initial learning rates were set to $0.0004$ and $0.0005$ for Adam and RMSProp, respectively. 
Also, the learning rate decay, dropping the learning rate linearly starting from epoch 10, is applied to the Adam optimizer. The parameter clipping was done for the critics and the range was set to $[-0.02, 0.02]$, and the number of training iterations for the critics was 5. The total number of training epochs was 30 and the mini-batch size was 64. 

\subsubsection{Ablation study on $\mathcal{L}_{\text{cyc}}$}

\begin{figure}[h]\vspace{-.1in}
\centering  
\subfigure[Iterative G2G (PSNR)]
{\includegraphics[width=0.4\linewidth]{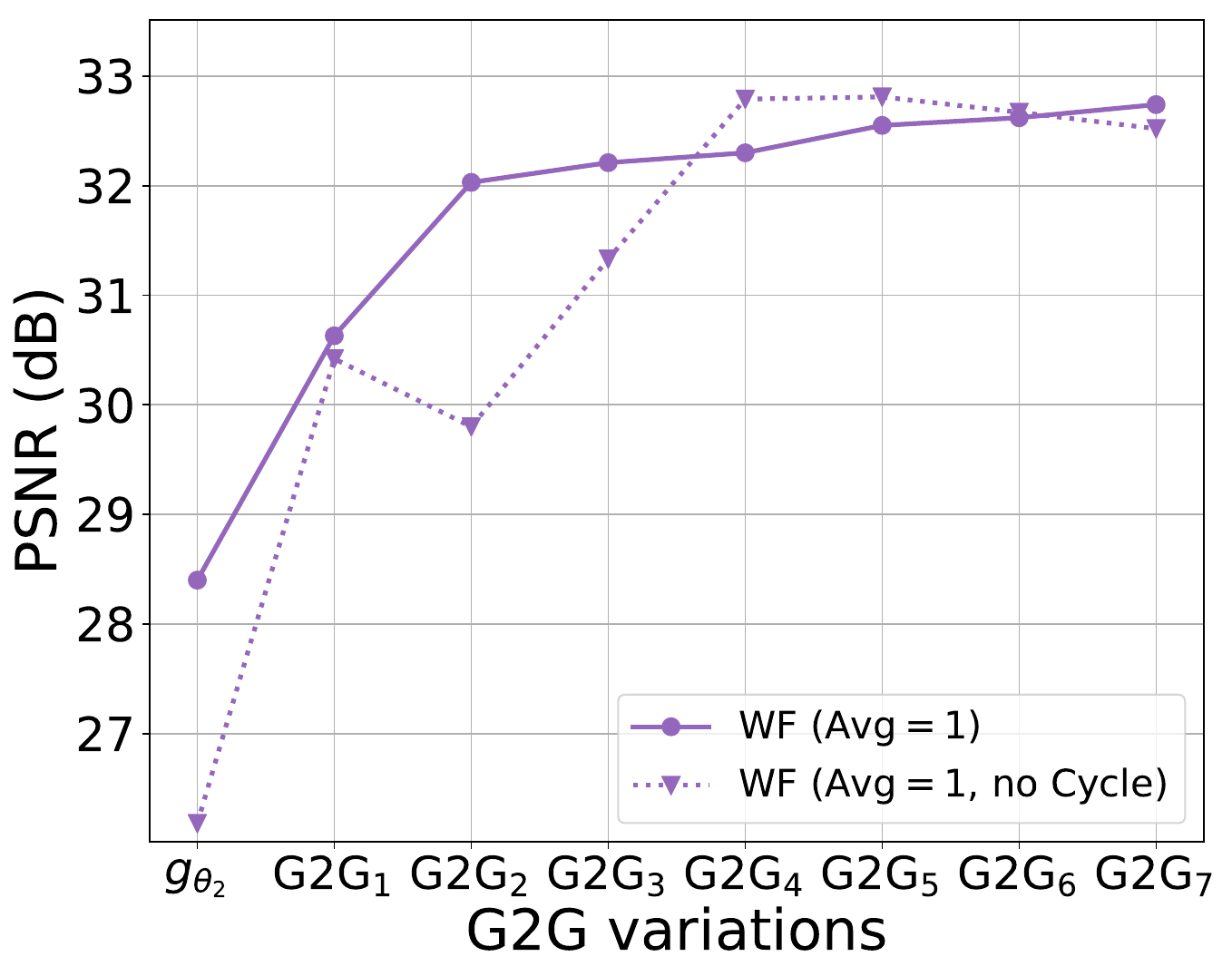}
\label{figure:ablation_cycle_psnr}}
\subfigure[Iterative G2G (SSIM)]
{\includegraphics[width=0.4\linewidth]{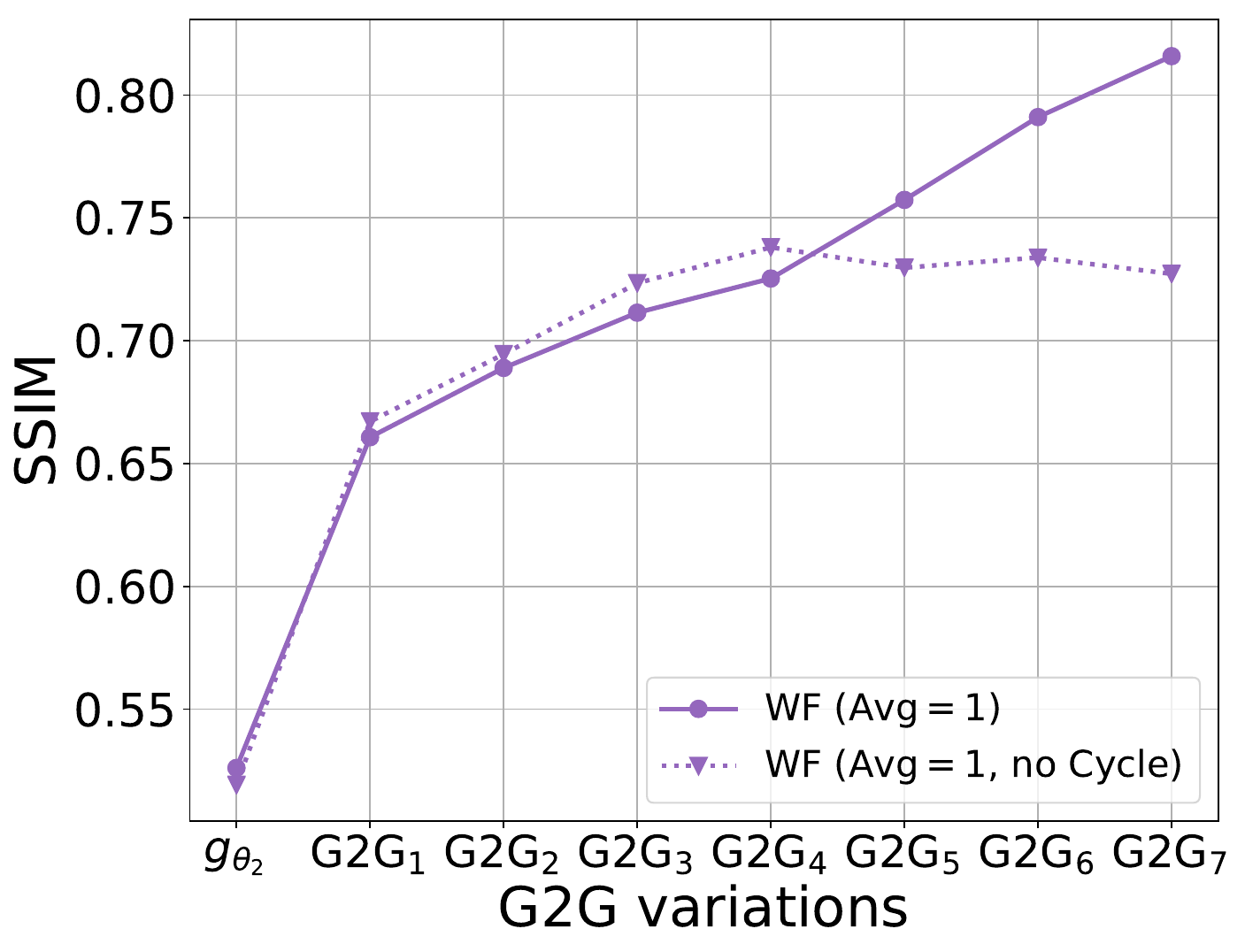}
\label{figure:ablation_cycle_ssim}}\vspace{-.1in}
\caption{Figure (a) and (b) compares the PSNR and SSIM performances between starting from $g_{\thetab_2}$ and $g_{\thetab_2}($No  $\mathcal{L}_{cyc})$, respectively.}\vspace{-.03in}
 \end{figure}


As shown in the synthetic noise case of Figure 6(b) (manuscript), the iterative G2G training is powerful such that there is a negligible performance difference between the schemes with and without $g_{\bm\theta_2}$, when the number of iterations is sufficiently large. Consequently, the cycle loss $\mathcal{L}_{cyc}$ also does not have significant effect in the final performance for the synthetic noise case. However, for the real noise case, $\mathcal{L}_{cyc}$ becomes more critical. As shown in Figure \ref{figure:ablation_cycle_psnr} and \ref{figure:ablation_cycle_ssim}, on WF(Avg$=1$) 
dataset, we observe that when there is no $\mathcal{L}_{cyc}$ in our generative model, the final PSNR or SSIM performances cannot reach the model \textit{with} $\mathcal{L}_{cyc}$ even after many iterations of G2G training. Hence, this result shows the necessity of $\mathcal{L}_{cyc}$.


\subsection{Iterative GAN2GAN training of a denoiser}
We do the same random cropping and data augmentation as in the generative model training. Moreover, for every minibatch in the G2G training, we generated new synthetic noisy image pairs using our trained generators as was done in the noise augmentation of \citep{ZhaZuoCheMenZha17}. 
Adam optimizer with
an initial learning rate $0.001$ was used, and the learning rate scheduling, which halves the learning rate every 20 epochs, was applied. The total number of training epochs was 50, and the mini-batch size was 4.
 We also stress that we set the architecture of $\hat{\Xb}_{\thetab}(\Zb)$ identical to that of 17-layers DnCNN in \citep{ZhaZuoCheMenZha17} to make a fair comparison. The pseudo algorithm for training a generative model is in Algorithm \ref{alg:g2g}
 
\begin{algorithm}[H]
        \caption{Training G2G, all experiments in this paper used the defaults values, $n_{epoch} = 50$, $\alpha_{G2G} = 1e^{-3}$}\label{alg:g2g}
        \begin{algorithmic}[1]
            \STATE \textbf{Require} $\mathcal{D}$, $g_{\theta_{1}}$, $g_{\theta_{2}}$, $\phi$, num\_iter, m
            \FOR{$j\gets 1, $ num\_iter}
                \FOR{$ep\gets 1, n_{epoch}$}
                    \STATE Sample $\{r^{(i)}_{j,1}, r^{(i)}_{j,2}\}^{m}_{i=1} \sim N(0,I),\, \{Z^{(i)}\}^{m}_{i=1} \sim \mathcal{D}$
                        \STATE $\bm\phi_j \gets \arg\min_{\bm\phi}\mathcal{L}_{\text{G2G}}(\bm\phi,\hat{\mathcal{D}}_j),$
                \ENDFOR
            \ENDFOR
            \STATE \textbf{return} $\bm\phi_{num_iter}$
        \end{algorithmic}
\end{algorithm}

\subsection{Noise2Void \citep{krull2018noise2void}}

We used the publicly available source code of Noise2Void (N2V) \citep{krull2018noise2void} to obtain the denoising results of N2V. Most of the hyperparameters were set to the default ones, but we changed three things to make a fair comparison with our method. 

Firstly, while the CNN architecture for the original N2V was a UNet3, we used the DnCNN \citep{zhang2018residual} with 17 layers such that it has the same structure as our G2G model. Secondly, as also is done in \citep{krull2018noise2void}, we had to use a validation set to do a proper model selection for N2V (i.e., the best epoch), while our G2G does not require any validation set (since we always use a model at the last epoch). The reason why N2V needs a validation is that its learning curve is very unstable and a proper model selection greatly affects the final denoising performance. To that end, since we used 20,500 patches with $120\times120$ size for training our G2G and other baselines, we divided the 20,500 patches into 18,000 training patches and 2,500 validation patches for training and selecting the best N2V model. Thirdly, we set 'mini\_batch\_size' to 4 (as our G2G) and 'train\_steps\_per\_epoch' to 'num\_of\_training\_data / mini\_batch\_size', hence, 4,500. Other hyperparameters are given in Table \ref{table:n2v_hyperparameter}.

\begin{table}[h]\caption{Hyperparameters for N2V}

\centering
\smallskip\noindent
\resizebox{.5\linewidth}{!}{
\begin{tabular}{|c|c|}
\hline
Hyperparameter             & Value             \\ \hline \hline
train\_steps\_per\_epoch   & 4,500              \\ \hline
train\_loss                & 'mse'             \\ \hline
train\_scheme              & 'Noise2Void'       \\ \hline
train\_batch\_size         & 4                 \\ \hline
n2v\_num\_pix              & 64                \\ \hline
n2v\_patch\_shape          & (64,64)           \\ \hline
n2v\_manupulator           & 'uniform\_withCP' \\ \hline
n2v\_neighborhood\_radious & '5'               \\ \hline
\end{tabular}}
    \label{table:n2v_hyperparameter}
\end{table}

\section{Comparison of the patch extraction rules}

\begin{figure*}[h]
\centering  
\subfigure[Example patch determined as homogeneous by the both rules.]{\includegraphics[width=0.8\linewidth]{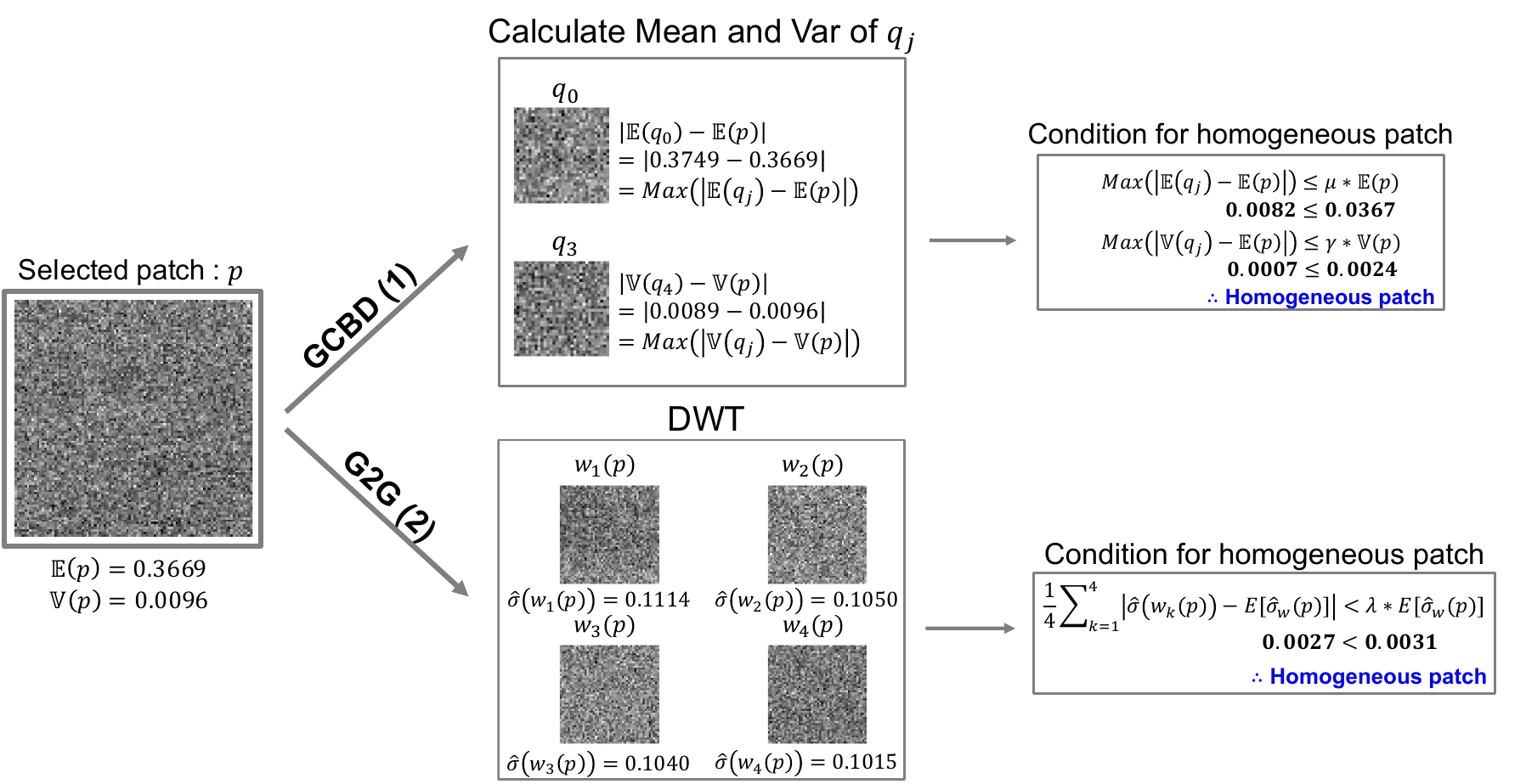}\label{figure:homogeneous_g2g_success}}
\subfigure[Example patch 1 that is wrongly determined as homogeneous by the GCBD rule]{\includegraphics[width=0.8\linewidth]{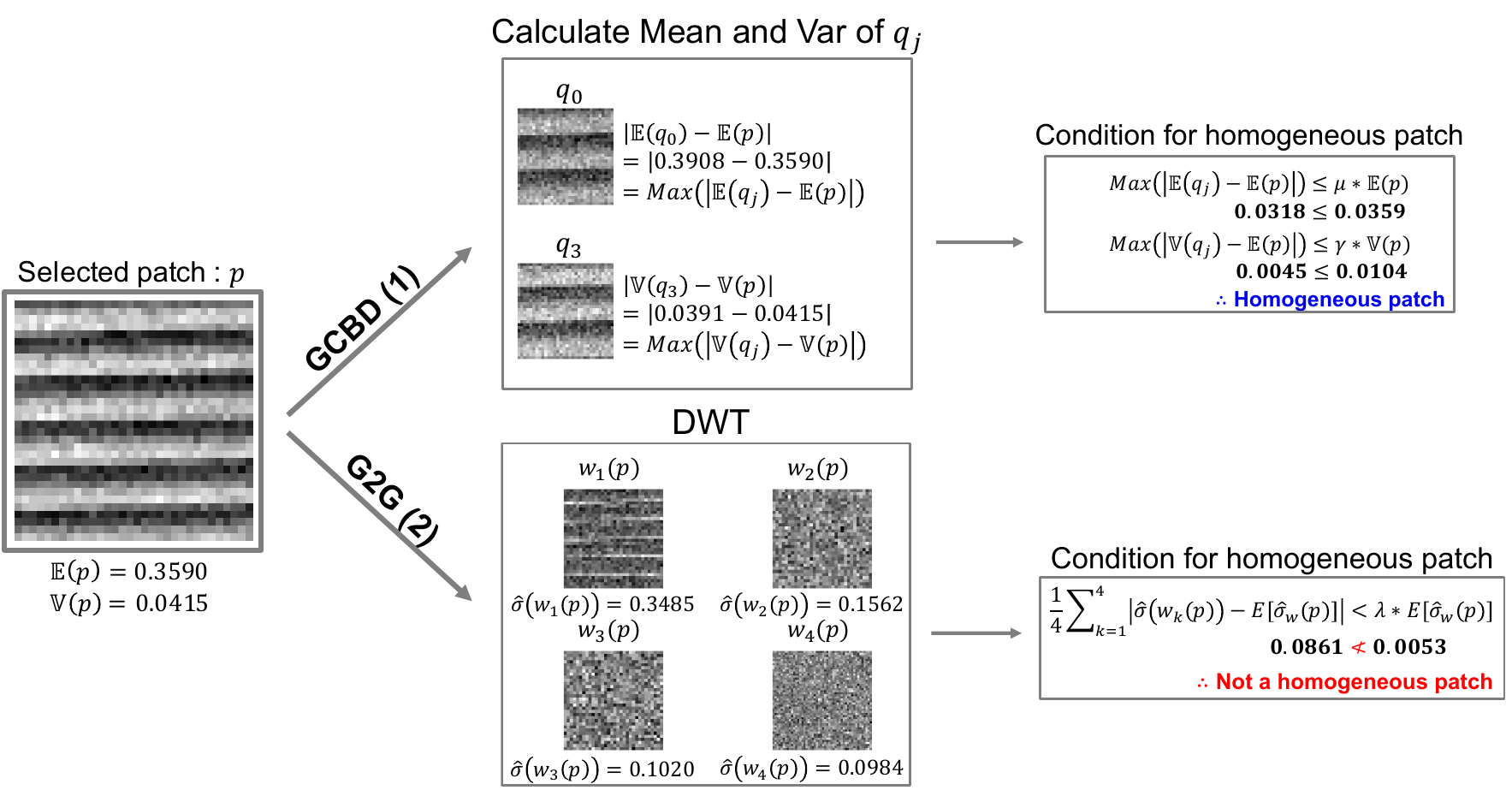}\label{figure:homogeneous_g2g_failure1}}
\subfigure[Example patch 2 that is wrongly determined as homogeneous by the GCBD rule]{\includegraphics[width=0.8\linewidth]{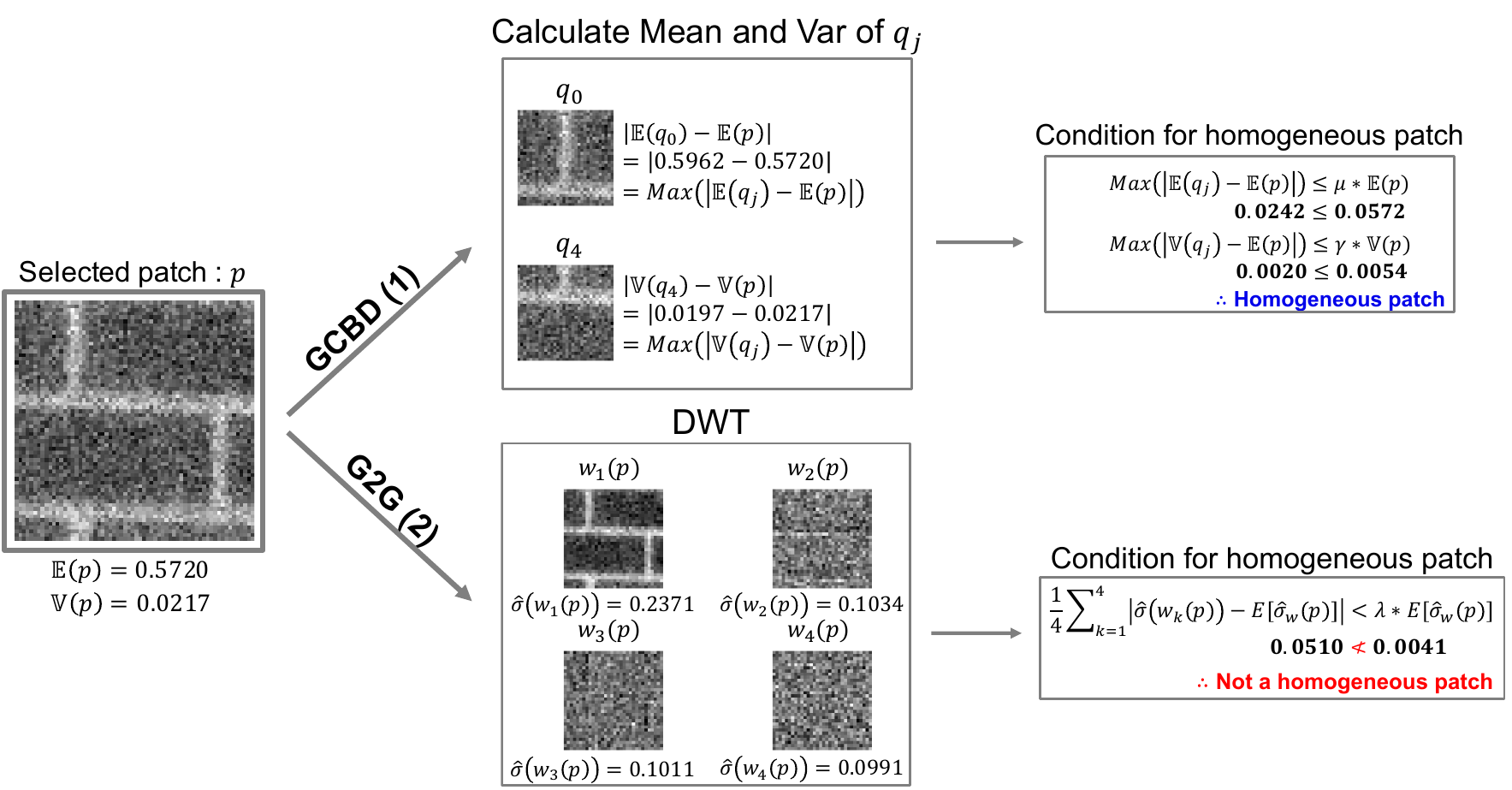}\label{figure:homogeneous_g2g_failure2}}
\caption{Noise patch extractions of GCBD (3) and G2G (4) rules.}\label{figure:homogeneous_g2g}
\vspace{-.1in}
\end{figure*}

Here, we make a further, thorough comparison between the GCBD smooth patch extraction rule [Eq.(3), manuscript] and ours [Eq.(4), manuscript]. We selected three noisy patches from [Figure 2(b), manuscript] and show the decision criterion of each rule for each image Figure \ref{figure:homogeneous_g2g}. From the figure, we can observe that while our G2G rule correctly excludes the patches in Figure \ref{figure:homogeneous_g2g_failure1} and \ref{figure:homogeneous_g2g_failure2} as non-homogeneous patches, the GCBD rule wrongly determines them also as homogeneous patches. That is, we note that since the DWT transform used in our rule can successfully disaggregate the high and low frequency components in the patches, the patches with \emph{self-similar repeating patterns} would have significantly varying sub-band coefficient variances as shown in the figures. Hence, our rule can exclude those patches.  However, in the GCBD rule, there may exist a sub-patch $\bm q_j$ that has similar empirical mean and variance as the original patch $\bm p$, thus, it may determine the patches with the self-similar repeating patterns as homogeneous as well. We believe these examples clearly show the stark difference between our rule and the GBCD rule for smooth patch extraction.

\clearpage

\subsection{Result table for real microscopy noise}
We report the detailed experimental results on the real microscopy images in Table \ref{table:real_microscopy}. We observe that Iterative G2G increases PSNR/SSIM in WF.

\begin{table}[h]\caption{Experimental results on the real microscopy dataset}

\centering
\smallskip\noindent
\resizebox{.98\linewidth}{!}{
\begin{tabular}{|c|c||c|c||c|c||c|c|c|c|c|c|c|c||c|c|}
\hline
\begin{tabular}[c]{@{}c@{}}Data\\ Type\end{tabular} & \begin{tabular}[c]{@{}c@{}}Noise\\ Type\end{tabular} & DnCNN-S                                                 & DnCNN$_B$                                               & \multicolumn{1}{l|}{BM3D}                                        & \begin{tabular}[c]{@{}c@{}}N2V\\ (DnCNN)\end{tabular}            & $g_{\theta_2}$                                                   & G2G$_1$                                                          & G2G$_2$                                                          & G2G$_3$                                                          & G2G$_4$                                                          & G2G$_5$                                                          & G2G$_6$                                                          & G2G$_7$                                                          & \begin{tabular}[c]{@{}c@{}}N2C\\ (GCBD)\end{tabular}    & \begin{tabular}[c]{@{}c@{}}N2C\\ ((Eq.(4))\end{tabular} \\ \hline \hline
\multirow{5}[12]{*}{WF}                                 & Raw                                                  & \begin{tabular}[c]{@{}c@{}}35.39\\ /0.8738\end{tabular} & \begin{tabular}[c]{@{}c@{}}25.43\\ /0.3702\end{tabular} & \begin{tabular}[c]{@{}c@{}}26.32\\ /0.4012\end{tabular}          & \begin{tabular}[c]{@{}c@{}}25.31\\ /0.3411\end{tabular}          & \begin{tabular}[c]{@{}c@{}}28.40\\ /0.5261\end{tabular}          & \begin{tabular}[c]{@{}c@{}}30.63\\ /0.6407\end{tabular}          & \begin{tabular}[c]{@{}c@{}}32.03\\ /0.6889\end{tabular}          & \begin{tabular}[c]{@{}c@{}}32.21\\ /0.7114\end{tabular}          & \begin{tabular}[c]{@{}c@{}}32.30\\ /0.7253\end{tabular}          & \begin{tabular}[c]{@{}c@{}}32.56\\ /0.7672\end{tabular}          & \begin{tabular}[c]{@{}c@{}}32.62\\ /0.7910\end{tabular}          & \begin{tabular}[c]{@{}c@{}}32.74\\ /0.8158\end{tabular}          & \begin{tabular}[c]{@{}c@{}}31.16\\ /0.7493\end{tabular} & \begin{tabular}[c]{@{}c@{}}32.26\\ /0.8205\end{tabular} \\ \cline{2-16} 
                                                    & Avg = 2                                              & \begin{tabular}[c]{@{}c@{}}36.11\\ /0.8969\end{tabular} & \begin{tabular}[c]{@{}c@{}}28.36\\ /0.5292\end{tabular} & \begin{tabular}[c]{@{}c@{}}29.21\\ /0.5642\end{tabular}          & \begin{tabular}[c]{@{}c@{}}28.23\\ /0.4500\end{tabular}          & \begin{tabular}[c]{@{}c@{}}29.73\\ /0.5844\end{tabular}          & \begin{tabular}[c]{@{}c@{}}31.84\\ /0.6717\end{tabular}          & \begin{tabular}[c]{@{}c@{}}32.41\\ /0.6920\end{tabular}          & \begin{tabular}[c]{@{}c@{}}32.80\\ /0.7161\end{tabular}          & \begin{tabular}[c]{@{}c@{}}32.85\\ /0.7500\end{tabular}          & \begin{tabular}[c]{@{}c@{}}32.90\\ /0.7697\end{tabular}          & \begin{tabular}[c]{@{}c@{}}32.92\\ /0.7783\end{tabular}          & \begin{tabular}[c]{@{}c@{}}32.85\\ /0.7808\end{tabular}          & \begin{tabular}[c]{@{}c@{}}31.88\\ /0.7575\end{tabular} & \begin{tabular}[c]{@{}c@{}}33.23\\ /0.8218\end{tabular} \\ \cline{2-16} 
                                                    & Avg = 4                                              & \begin{tabular}[c]{@{}c@{}}37.46\\ /0.9182\end{tabular} & \begin{tabular}[c]{@{}c@{}}31.32\\ /0.6910\end{tabular} & \begin{tabular}[c]{@{}c@{}}32.19\\ /0.7202\end{tabular}          & \begin{tabular}[c]{@{}c@{}}31.28\\ /0.6676\end{tabular}          & \begin{tabular}[c]{@{}c@{}}31.41\\ /0.6580\end{tabular}          & \begin{tabular}[c]{@{}c@{}}33.32\\ /0.7728\end{tabular}          & \begin{tabular}[c]{@{}c@{}}33.52\\ /0.7974\end{tabular}          & \begin{tabular}[c]{@{}c@{}}33.71\\ /0.8079\end{tabular}          & \begin{tabular}[c]{@{}c@{}}33.68\\ /0.8091\end{tabular}          & \begin{tabular}[c]{@{}c@{}}33.85\\ /0.8140\end{tabular}          & \begin{tabular}[c]{@{}c@{}}33.69\\ /0.8088\end{tabular}          & \begin{tabular}[c]{@{}c@{}}33.79\\ /0.8135\end{tabular}          & \begin{tabular}[c]{@{}c@{}}34.42\\ /0.8665\end{tabular} & \begin{tabular}[c]{@{}c@{}}34.79\\ /0.8559\end{tabular} \\ \cline{2-16} 
                                                    & Avg = 8                                              & \begin{tabular}[c]{@{}c@{}}39.81\\ /0.9374\end{tabular} & \begin{tabular}[c]{@{}c@{}}34.63\\ /0.8218\end{tabular} & \begin{tabular}[c]{@{}c@{}}35.76\\ /0.8444\end{tabular}          & \begin{tabular}[c]{@{}c@{}}34.85\\ /0.8097\end{tabular}          & \begin{tabular}[c]{@{}c@{}}34.81\\ /0.8084\end{tabular}          & \begin{tabular}[c]{@{}c@{}}35.16\\ /0.8315\end{tabular}          & \begin{tabular}[c]{@{}c@{}}35.16\\ /0.8315\end{tabular}          & \begin{tabular}[c]{@{}c@{}}35.27\\ /0.8325\end{tabular}          & \begin{tabular}[c]{@{}c@{}}35.21\\ /0.8321\end{tabular}          & \begin{tabular}[c]{@{}c@{}}35.27\\ /0.8333\end{tabular}          & \begin{tabular}[c]{@{}c@{}}35.25\\ /0.8330\end{tabular}          & \begin{tabular}[c]{@{}c@{}}35.22\\ /0.8316\end{tabular}          & \begin{tabular}[c]{@{}c@{}}36.92\\ /0.9126\end{tabular} & \begin{tabular}[c]{@{}c@{}}36.86\\ /0.8800\end{tabular} \\ \cline{2-16} 
                                                    & Avg = 16                                             & \begin{tabular}[c]{@{}c@{}}42.10\\ /0.9569\end{tabular} & \begin{tabular}[c]{@{}c@{}}37.82\\ /0.9136\end{tabular} & \begin{tabular}[c]{@{}c@{}}39.67\\ /0.9293\end{tabular}          & \begin{tabular}[c]{@{}c@{}}38.75\\ /0.9094\end{tabular}          & \begin{tabular}[c]{@{}c@{}}36.97\\ /0.9086\end{tabular}          & \begin{tabular}[c]{@{}c@{}}38.97\\ /0.9153\end{tabular}          & \begin{tabular}[c]{@{}c@{}}38.98\\ /0.9174\end{tabular}          & \begin{tabular}[c]{@{}c@{}}38.84\\ /0.9172\end{tabular}          & \begin{tabular}[c]{@{}c@{}}38.84\\ /0.9170\end{tabular}          & \begin{tabular}[c]{@{}c@{}}38.87\\ /0.9175\end{tabular}          & \begin{tabular}[c]{@{}c@{}}38.82\\ /0.9181\end{tabular}          & \begin{tabular}[c]{@{}c@{}}38.82\\ /0.9178\end{tabular}          & \begin{tabular}[c]{@{}c@{}}38.72\\ /0.9110\end{tabular} & \begin{tabular}[c]{@{}c@{}}38.92\\ /0.9181\end{tabular} \\ \hline \hline
\multicolumn{2}{|c||}{Average}                                                                              & \begin{tabular}[c]{@{}c@{}}38.17\\ /0.9166\end{tabular} & \begin{tabular}[c]{@{}c@{}}31.52\\ /0.6652\end{tabular} & \textbf{\begin{tabular}[c]{@{}c@{}}32.63\\ /0.6919\end{tabular}} & \textbf{\begin{tabular}[c]{@{}c@{}}31.68\\ /0.6365\end{tabular}} & \textbf{\begin{tabular}[c]{@{}c@{}}32.26\\ /0.6971\end{tabular}} & \textbf{\begin{tabular}[c]{@{}c@{}}33.98\\ /0.7664\end{tabular}} & \textbf{\begin{tabular}[c]{@{}c@{}}34.41\\ /0.7855\end{tabular}} & \textbf{\begin{tabular}[c]{@{}c@{}}34.57\\ /0.7970\end{tabular}} & \textbf{\begin{tabular}[c]{@{}c@{}}34.58\\ /0.8067\end{tabular}} & \textbf{\begin{tabular}[c]{@{}c@{}}34.69\\ /0.8203\end{tabular}} & \textbf{\begin{tabular}[c]{@{}c@{}}34.66\\ /0.8258\end{tabular}} & \textbf{\begin{tabular}[c]{@{}c@{}}34.68\\ /0.8324\end{tabular}} & \begin{tabular}[c]{@{}c@{}}34.62\\ /0.8394\end{tabular} & \begin{tabular}[c]{@{}c@{}}35.21\\ /0.8592\end{tabular} \\ \hline
\end{tabular}}
    \label{table:real_microscopy}
\end{table}

\subsection{Description and the result table on Reconstructed CT dataset}

The reconstructed CT dataset consists of chest and head parts of 27 pediatric extended cardiac-torso phantoms \citep{segars2015development}, which provide a highly realistic model of the human anatomy. We extracted 60 image slices from each phantom, leading to 1620 image slices in total. The dataset was generated in the following procedure. First, noiseless projection data were acquired in a parallel-beam geometry with Siddon’s ray-driven algorithm \citep{sidky2008image}. To reduce view aliasing artifacts, the detector quarter-offset was used during a CT scan. Second, Poisson noise was generated and added to the noiseless projection data. Note that the mean number of detected photons was set to 2,500, 5,000, 7,500, and 10,000 to simulate 25\%, 50\%, 75\%, and 100\% of a normal dose, respectively. Finally, the images were reconstructed by filtered backprojection \citep{hsieh2003computed}. To preserve fine anatomical structures in the images, the Ram-Lak filter was used as a reconstruction filter. Detailed simulation parameters are summarized in Table \ref{table:description}.

\begin{table}[h]\caption{Simulation parameters}

\centering
\smallskip\noindent
\resizebox{.4\linewidth}{!}{
\begin{tabular}{|c|c|}
\hline
Parameters                    & Values    \\ \hline \hline
Source to iso-center distance & 595 mm    \\ \hline
Source to detector distance   & mm        \\ \hline
Detector cell size            & 0.7 mm    \\ \hline
Detector array size           & 736 x 1   \\ \hline
Data acquisition angle        & 360 dares \\ \hline
Number of projection views    & 736       \\ \hline
Reconstructed pixel width     & 0.67 mm   \\ \hline
Reconstructed matrix size     & 512x512   \\ \hline
\end{tabular}}
    \label{table:description}
\end{table}

\begin{figure}[h!]
\centering  
\includegraphics[width=0.9\linewidth]{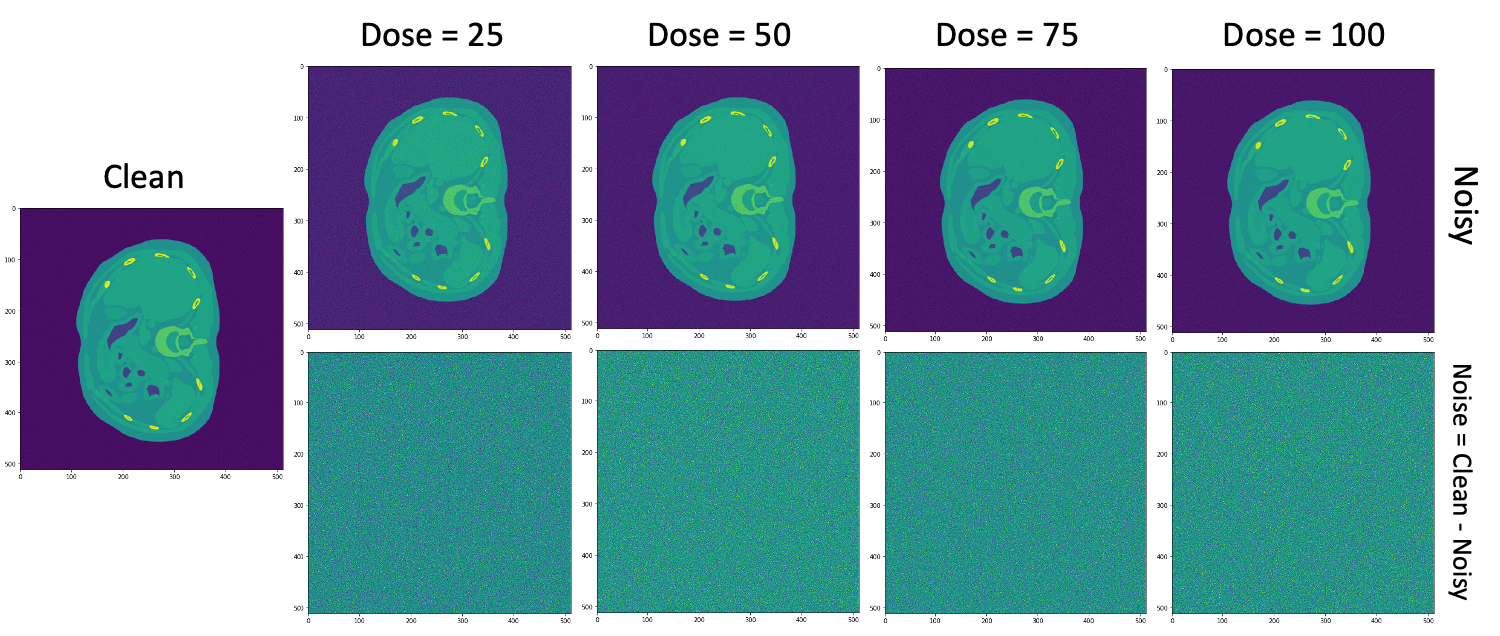}
\caption{Clean, noisy and noise images from Reconstructed CT}\label{figure:vis_medical}
\end{figure}

We divided 27 phantoms into training and test data and the phantom number for each dataset is in Fig \ref{table:tr_te_info}. Also, We visualized the first image of Female 1 in Fig \ref{figure:vis_medical}. We can clearly see that each dose has a different noise level, and the noise is source independent and correlated. Finally, Table \ref{table:medical} shows the details of experimental results on Reconstructed CT dataset.

\begin{table}[h]\caption{Training and test data information of Reconstructed CT dataset}

\centering
\smallskip\noindent
\resizebox{.5\linewidth}{!}{
\begin{tabular}{|c|c|c|}
\hline
             & Training data           & Test data  \\ \hline \hline
Female       & 1,3,4,5,6,7,8,9,10,11   & 13,14,15   \\ \hline
Male         & 1,2,3,4,5,6,7,8,9,10,11 & 12,13      \\ \hline \hline
\# of images & 21x60 = 1260            & 60x5 = 300 \\ \hline
\end{tabular}}
    \label{table:tr_te_info}
\end{table}

\newpage

\begin{table}[h]\caption{Experimental results on the reconstructed CT}

\centering
\smallskip\noindent
\resizebox{.98\linewidth}{!}{
\begin{tabular}{|c|c||c|c||c|c||c|c|c|c|}
\hline
\begin{tabular}[c]{@{}c@{}}Data\\ Type\end{tabular}                         & \begin{tabular}[c]{@{}c@{}}Noise\\ Type\end{tabular} & \begin{tabular}[c]{@{}c@{}}N2C\\ (UNnet)\end{tabular}   & DnCNN$_B$                                               & \multicolumn{1}{l|}{BM3D}                                        & \begin{tabular}[c]{@{}c@{}}N2V\\ (UNet)\end{tabular}             & $g_{\theta_2}$                                                   & G2G$_1$                                                          & G2G$_2$                                                          & G2G$_3$                                                          \\ \hline \hline
\multirow{4}[10]{*}{\begin{tabular}[c]{@{}c@{}}Reconstructed\\ CT\end{tabular}} & Dose 25                                              & \begin{tabular}[c]{@{}c@{}}48.43\\ /0.9609\end{tabular} & \begin{tabular}[c]{@{}c@{}}35.50\\ /0.6055\end{tabular} & \begin{tabular}[c]{@{}c@{}}42.40\\ /0.7575\end{tabular}          & \begin{tabular}[c]{@{}c@{}}31.57\\ /0.5416\end{tabular}          & \begin{tabular}[c]{@{}c@{}}34.66\\ /0.5759\end{tabular}          & \begin{tabular}[c]{@{}c@{}}40.49\\ /0.8301\end{tabular}          & \begin{tabular}[c]{@{}c@{}}46.04\\ /0.9579\end{tabular}          & \begin{tabular}[c]{@{}c@{}}47.47\\ /0.9707\end{tabular}          \\ \cline{2-10} 
                                                                            & Dose 50                                              & \begin{tabular}[c]{@{}c@{}}49.07\\ /0.9600\end{tabular} & \begin{tabular}[c]{@{}c@{}}38.48\\ /0.7440\end{tabular} & \begin{tabular}[c]{@{}c@{}}45.14\\ /0.8510\end{tabular}          & \begin{tabular}[c]{@{}c@{}}34.17\\ /0.6931\end{tabular}          & \begin{tabular}[c]{@{}c@{}}37.70\\ /0.7202\end{tabular}          & \begin{tabular}[c]{@{}c@{}}43.38\\ /0.9063\end{tabular}          & \begin{tabular}[c]{@{}c@{}}47.78\\ /0.9702\end{tabular}          & \begin{tabular}[c]{@{}c@{}}48.06\\ /0.9705\end{tabular}          \\ \cline{2-10} 
                                                                            & Dose 75                                              & \begin{tabular}[c]{@{}c@{}}49.45\\ /0.9591\end{tabular} & \begin{tabular}[c]{@{}c@{}}40.09\\ /0.8111\end{tabular} & \begin{tabular}[c]{@{}c@{}}46.55\\ /0.8929\end{tabular}          & \begin{tabular}[c]{@{}c@{}}35.90\\ /0.7756\end{tabular}          & \begin{tabular}[c]{@{}c@{}}39.49\\ /0.7919\end{tabular}          & \begin{tabular}[c]{@{}c@{}}44.96\\ /0.9320\end{tabular}          & \begin{tabular}[c]{@{}c@{}}48.80\\ /0.9744\end{tabular}          & \begin{tabular}[c]{@{}c@{}}49.20\\ /0.9733\end{tabular}          \\ \cline{2-10} 
                                                                            & Dose 100                                             & \begin{tabular}[c]{@{}c@{}}49.63\\ /0.9565\end{tabular} & \begin{tabular}[c]{@{}c@{}}41.19\\ /0.8513\end{tabular} & \begin{tabular}[c]{@{}c@{}}47.48\\ /0.9169\end{tabular}          & \begin{tabular}[c]{@{}c@{}}37.26\\ /0.8118\end{tabular}          & \begin{tabular}[c]{@{}c@{}}40.75\\ /0.9350\end{tabular}          & \begin{tabular}[c]{@{}c@{}}46.11\\ /0.9492\end{tabular}          & \begin{tabular}[c]{@{}c@{}}49.19\\ /0.9760\end{tabular}          & \begin{tabular}[c]{@{}c@{}}48.83\\ /0.9718\end{tabular}          \\ \hline \hline
\multicolumn{2}{|c||}{Average}                                                                                                      & \begin{tabular}[c]{@{}c@{}}49.15\\ /0.9591\end{tabular} & \begin{tabular}[c]{@{}c@{}}38.82\\ /0.5730\end{tabular} & \textbf{\begin{tabular}[c]{@{}c@{}}45.39\\ /0.8546\end{tabular}} & \textbf{\begin{tabular}[c]{@{}c@{}}34.73\\ /0.7055\end{tabular}} & \textbf{\begin{tabular}[c]{@{}c@{}}38.15\\ /0.7558\end{tabular}} & \textbf{\begin{tabular}[c]{@{}c@{}}43.74\\ /0.9044\end{tabular}} & \textbf{\begin{tabular}[c]{@{}c@{}}47.95\\ /0.9696\end{tabular}} & \textbf{\begin{tabular}[c]{@{}c@{}}48.39\\ /0.9715\end{tabular}} \\ \hline
\end{tabular}}
    \label{table:medical}
\end{table}

\section{Analysis on real microscopy image}
In this section, we analyze why our G2G also works well on the real microscopy image dataset (WF) \cite{zhang2018poisson}. although the source-dependent noise does not satisfy our assumption on the noise.
The real microscopy image dataset consists of three different types of dataset, which are Wide-Focal(WF), Two-Photon(TP) and Con-Focal(CF), and 
It is generally known that the real noise follows the Poisson-Gaussian model \cite{zhang2018poisson},
\begin{equation}
Z_{i} = x_i+N_{i}, \ \ \ i = 1,2,\ldots,
\end{equation}
in which $N_i\sim\mathcal{N}(0,\sigma_i^2)$ and 
\begin{equation}
    \sigma_{i}^2= \alpha x_{i} + \sigma^2\label{sigma}
\end{equation}
with a scaling factor $\alpha>0$. Thus, the noise variance depends on the underlying clean source pixel value, and $\alpha$ determines the level of the dependence. 





\begin{figure}[h]
\centering  
\includegraphics[width=0.9\linewidth]{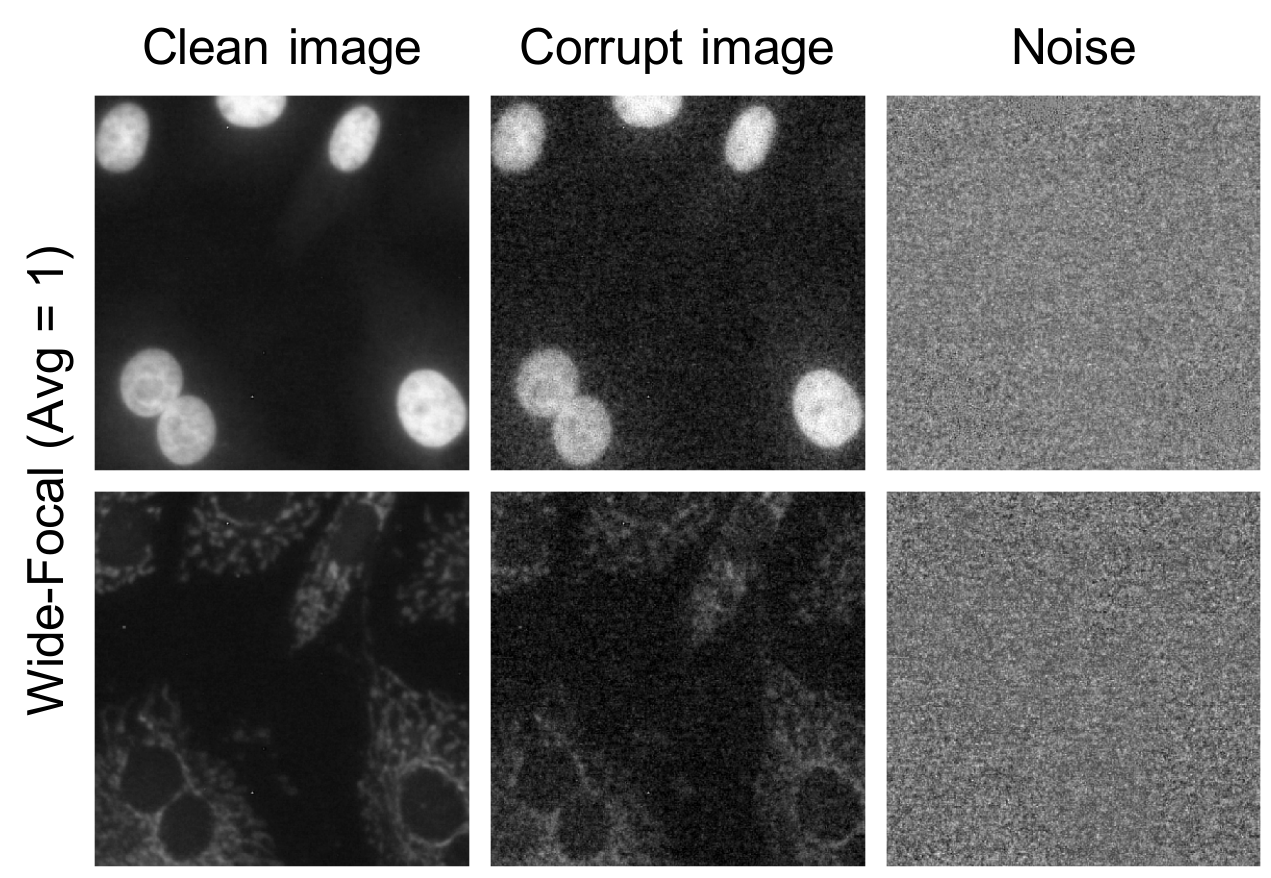}
\caption{Clean, noisy and noise images from the WF set. (Best viewed in PDF.)}\label{figure:wide_focal}
\end{figure} 

In Table \ref{table:real_microscopy}, we observe that our G2G performs well for WF compared to other baselines, 
hence, we visualize clean, noisy, noise images from each set and examine if there are any notable difference in the noise distributions. 
Figure \ref{figure:wide_focal} shows two image samples (Avg$=1$ cases) from the Wide-Focal (WF) set. The noise images are obtained by subtracting the clean images from its noisy versions. 
Note even though the intensities in the source images change significantly among pixels (particularly for the top image), the noise images do not show any source-dependent patterns. Hence, we can deduce that $\alpha$ may be small for the WF images. Also, we could see that there is a correlated pattern in the noise. We belive that these back the good performance of G2G for the WF set. 


\begin{figure}[h]
\centering  
\includegraphics[width=0.9\linewidth]{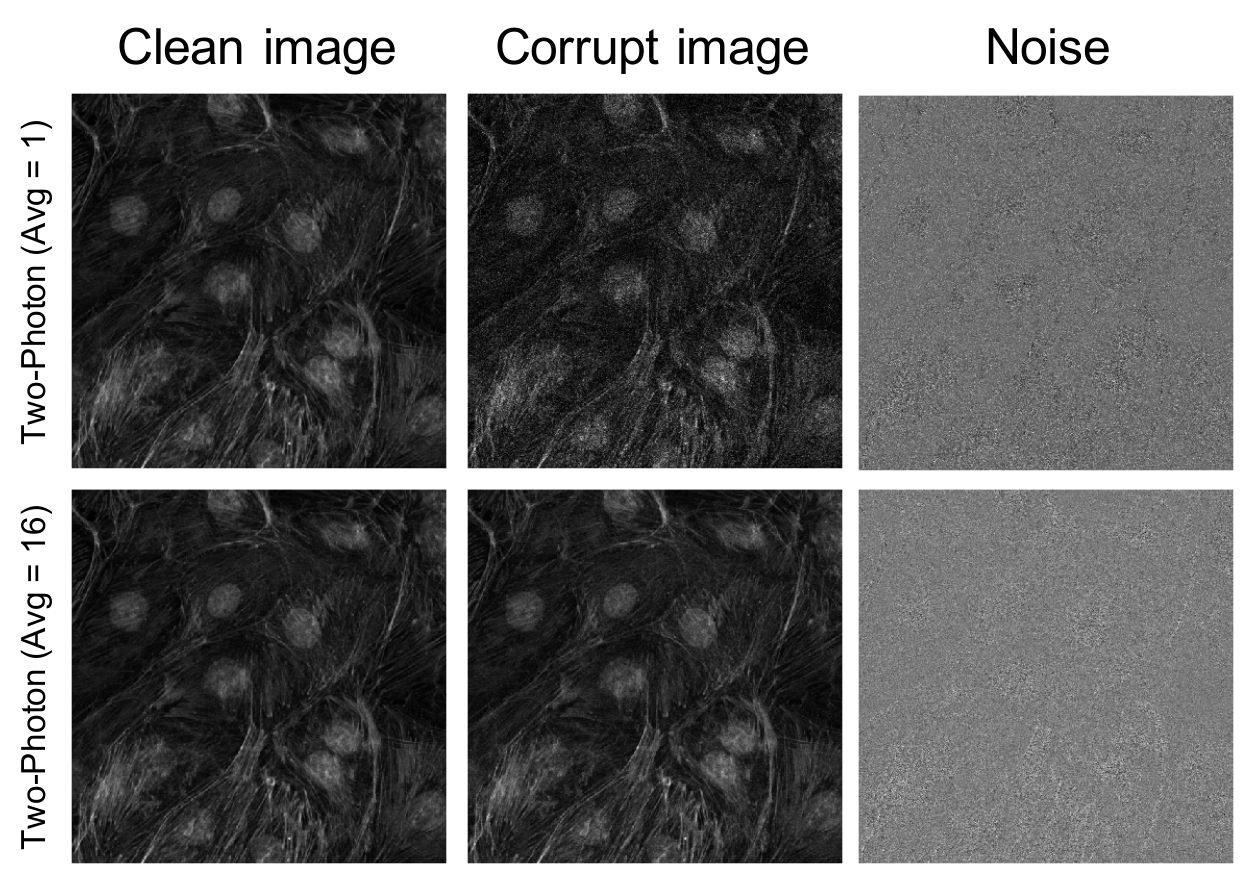}
\caption{Clean, noisy and noise images from the TP set. (Best viewed in PDF.)}\label{figure:two_photon}
\end{figure} 

Figure \ref{figure:two_photon}, on the other hand, visualizes an image from TP set for Avg$=\{1, 16\}$ cases. Comparing with Figure \ref{figure:wide_focal}, we can clearly see the source-dependent patterns in the noise images, particularly severely for the $\text{Avg}=1$ case. Also, CP set showed the similar source-dependent noise patterns. This source-dependent noise is not in our assumption so we did not apply GAN2GAN to TP and CP. However, we want to stress out that the source independent real noise also exists and GAN2GAN shows the best result compared to any other baselines.

\section{Visualizations}

\subsection{Visualization of $\tilde{\Zb}$}

\begin{figure*}[h]
\centering  
\includegraphics[width=0.95\linewidth]{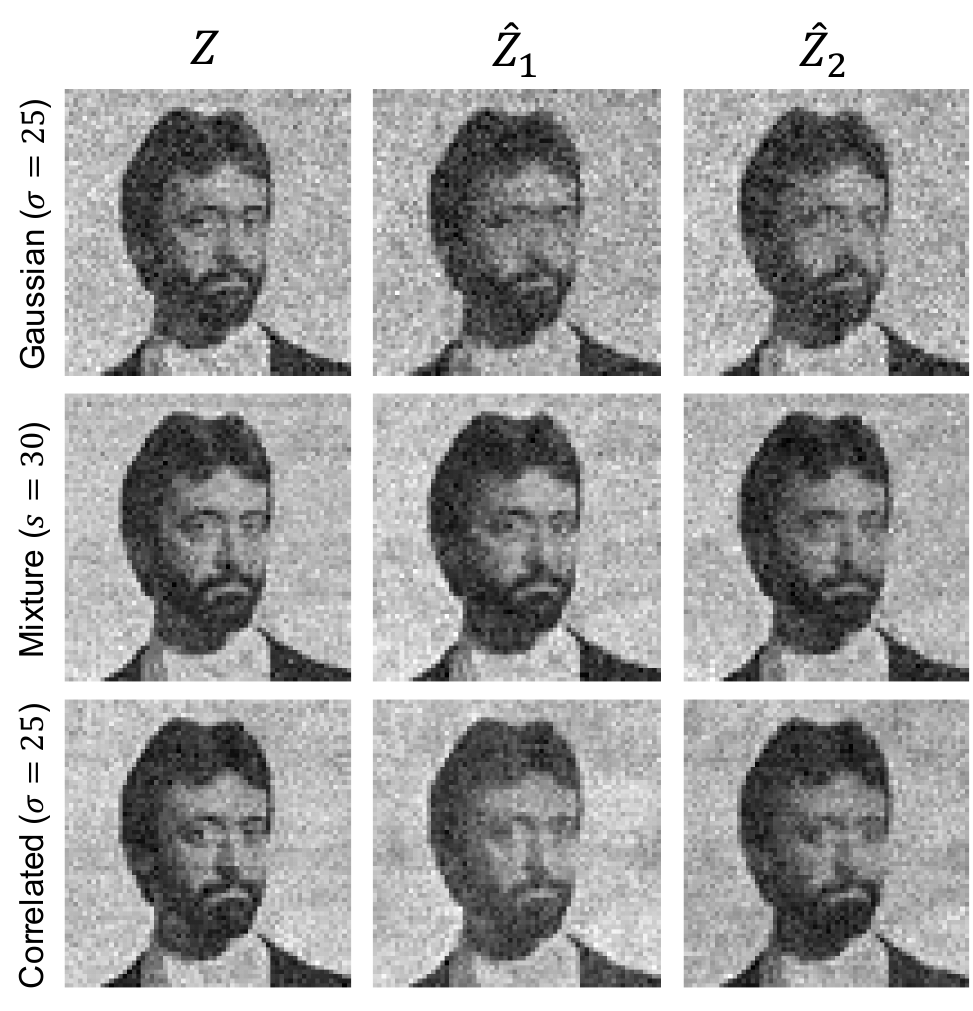}
\caption{Visualizations of synthesized synthetic noisy image pairs.}
\label{figure:z_tilde_synthetic}
\end{figure*}

\begin{figure*}[h]
\centering  
\includegraphics[width=0.9\linewidth]{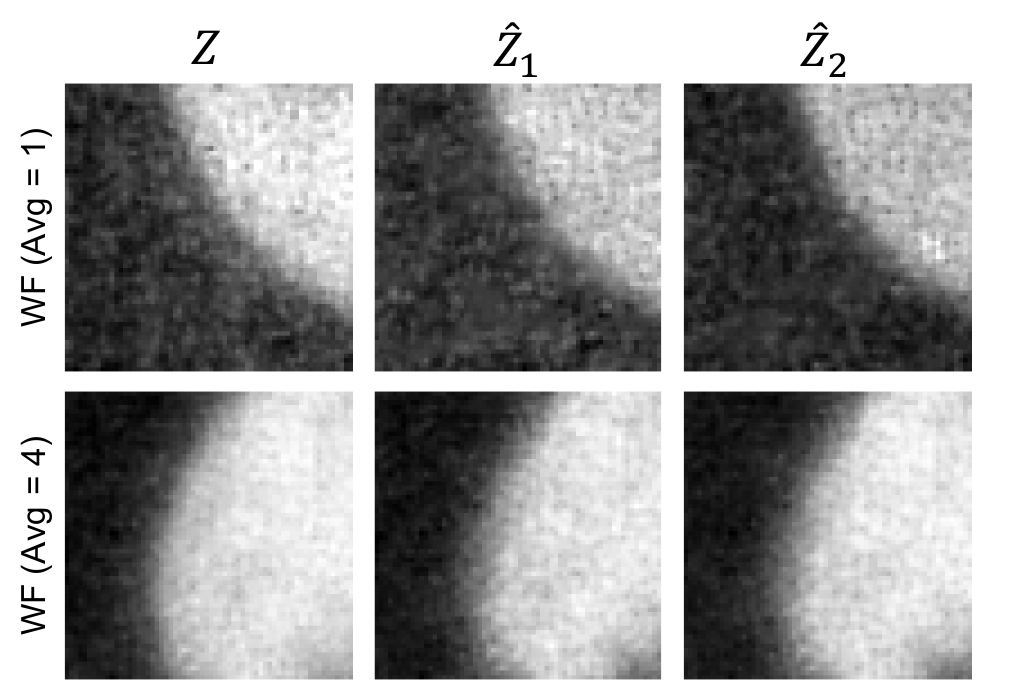}
\caption{Visualizations of synthesized real noise image pairs.}
\label{figure:z_tilde_real}
\end{figure*}

Figure \ref{figure:z_tilde_synthetic} and \ref{figure:z_tilde_real} visualize the simulated noisy image pairs $(\hat{\mathbf{Z}}_1, \hat{\mathbf{Z}}_2)$, generated from our generative model, for synthetic and real noise cases, respectively. A close examination shows that the images are not simple copies of the original noisy image $\mathbf{Z}$ but are successfully synthesized with the independent noise processes. 

\clearpage

\subsection{Visualization of denoised images on BSD68}

\begin{figure*}[h]
\centering  
\includegraphics[width=0.98\linewidth]{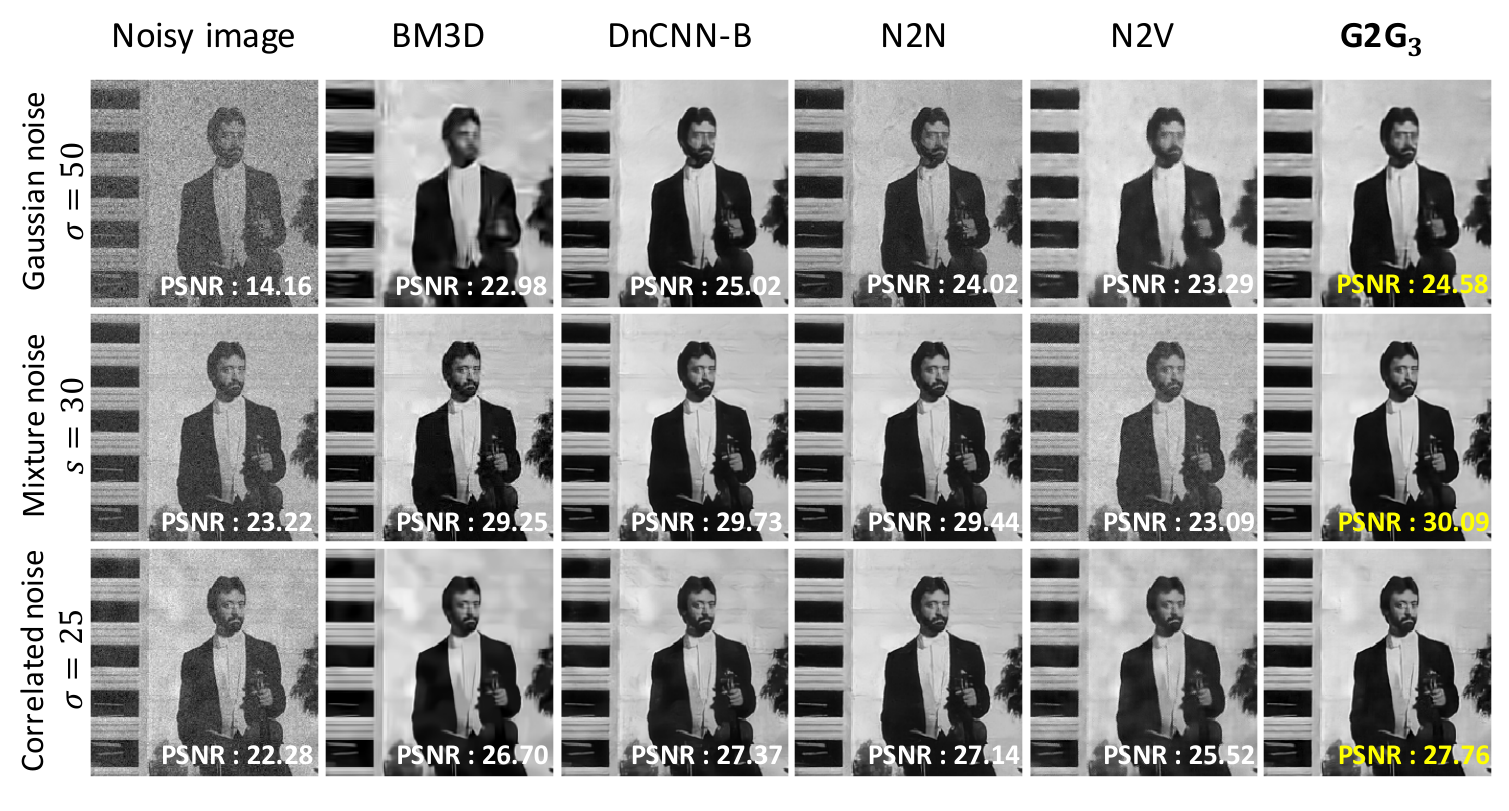}
\caption{Denoising results on the synthetic noise images.}
\label{figure:visualizations_BSD}
\end{figure*}

Figure \ref{figure:visualizations_BSD} visualizes the denoising results of a BSD68 image for different types of noise. Note the clear difference in the noise characteristics for Gaussian, mixture, and correlated noises. The visualization of G2G$_{3}$ certiainly seems better than N2V and BM3D, in line with the PSNR results. DnCNN-B and Noise2Noise use more information than G2G$_{3}$, but the visualzation as well as the PSNR of G2G$_{3}$ are comparable to those of the two methods.  

\clearpage


\subsection{Additional visualizations on the real microscopy images}

\begin{figure*}[h]
\centering  
\includegraphics[width=0.75\linewidth]{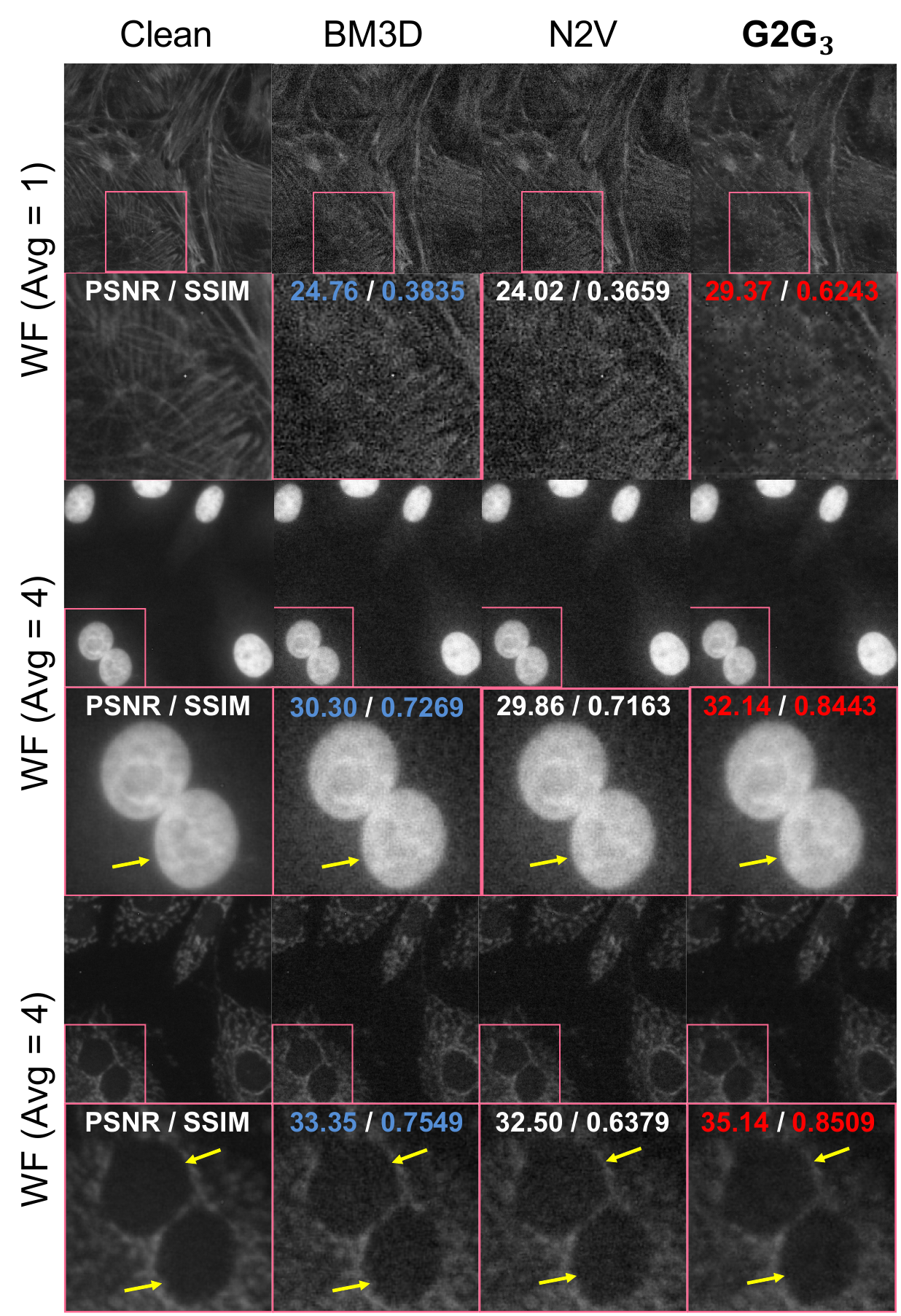}
\caption{Denoising results on the real noisy microscopy images.}
\label{figure:visualizations_WFTP}
\end{figure*}

We also visualize additional denoised images of WF images in Figure \ref{figure:visualizations_WFTP}. We can see that the denoising results of the baselines for WF (Avg = 1) are very noisy, but G2G$_{3}$ shows relatively clean denoising results than others.

\clearpage

\subsection{Additional visualizations on the Reconstructed CT}

\begin{figure*}[h]
\centering  
\includegraphics[width=0.98\linewidth]{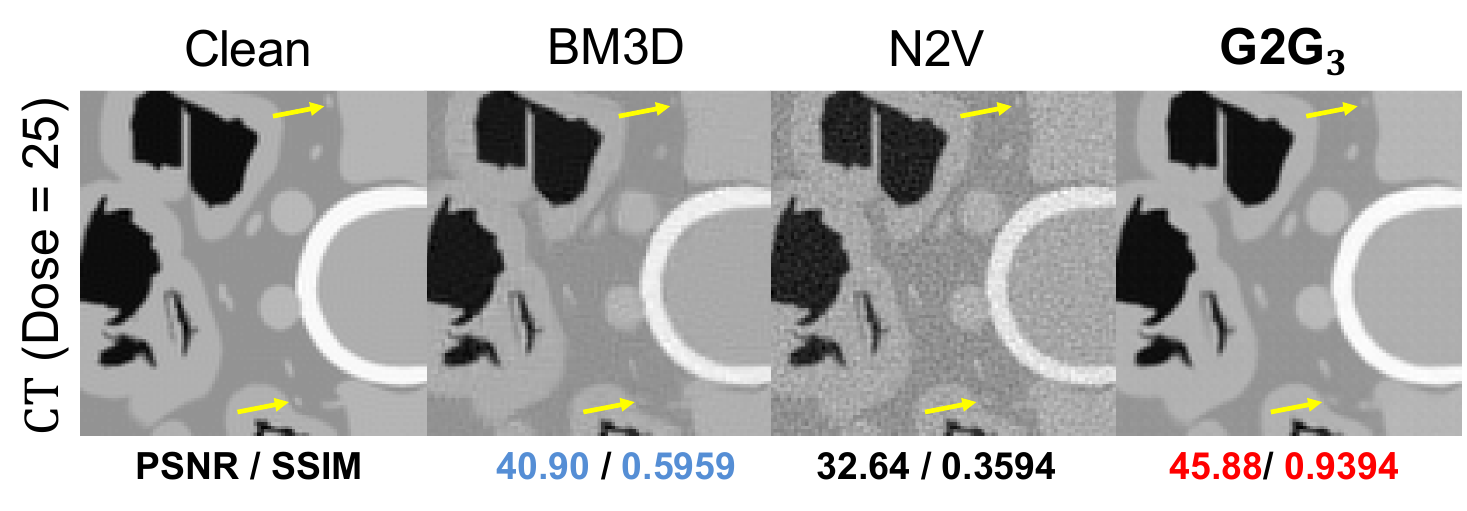}
\caption{Denoising results on Reconstructed CT images.}
\label{figure:visualizations_medical}
\end{figure*}

We visualize the denoising result of a Reconstructed CT image in Figure \ref{figure:visualizations_medical}. We observed that BM3D and N2V shows a stil noisy result on this image but G2G$_3$ shows a clearly denoised result.

\section{Experimental Results of G2G$_3$ with $(\hat{\Zb}_{j1}^{(i)}, \hat{X}_{\bm\phi_{j-1}}(\Zb^{(i)}))$}

\begin{table}[h]\caption{Experimental results for Reviewer 3's Q.(2).}

\centering
\smallskip\noindent
\resizebox{.6\linewidth}{!}{
\begin{tabular}{|c||c|c|}
\hline
\multirow{2}[2]{*}{PSNR / SSIM} & \multicolumn{2}{c|}{G2G$_3$}                                                                                 \\ \cline{2-3} 
                             & $(\hat{\Zb}_{j1}^{(i)}, \hat{X}_{\bm\phi_{j-1}}(\Zb^{(i)}))$ & $(\hat{\Zb}_{j1}^{(i)},\hat{\Zb}_{j2}^{(i)})$ \\ \hline \hline
Gaussian ($\sigma = 25$)     & 29.02 / 0.8153                                               & 28.96 / 0.8080                                \\ \hline
Mixture ($s = 30$)           & 30.70 / 0.8621                                               & 30.49 / 0.8538                                \\ \hline
Correlated ($\sigma = 25$)   & 24.04 / 0.8673                                               & 28.00 / 0.8447                                \\ \hline
WF (Avg = 1)                 & 16.24 / 0.4490                                               & 34.57 / 0.7970                                \\ \hline
Medical (Dose = 25)          & 18.52 / 0.7397                                               & 47.47 / 0.9707                                \\ \hline
\end{tabular}}
    \label{table:g2g_x_hat}
\end{table}

 \textcolor{black}{We did the experiments on G2G$_3$ with $(\hat{\Zb}_{j1}^{(i)}, \hat{X}_{\bm\phi_{j-1}}(\Zb^{(i)}))$ and Table \ref{table:g2g_x_hat} shows the experimental results. From the table, we observe the suggested approach (left column) can in fact achieve slightly improved results for synthetic Gaussian and Mixture noise. However, we observe that the performances of the approach for the Correlated and WF/Medical datasets deteriorate significantly compared to ours.}

\clearpage



\bibliographystyle{iclr2021_conference}
\bibliography{bibfile.bib}